\documentclass[10pt]{bmc_article}

\usepackage{cite} % Make references as [1-4], not [1,2,3,4]
\usepackage{url}  % Formatting web addresses  
\usepackage{ifthen}  % Conditional 
\usepackage{multicol}   %Columns
\usepackage[utf8]{inputenc} %unicode support
 \usepackage{amsmath}
\usepackage{graphicx}
\usepackage{color}
 
\newboolean{publ}

\newenvironment{bmcformat}{\baselineskip20pt\sloppy\setboolean{publ}{false}}{\baselineskip20pt\sloppy}

\begin{document}
\begin{bmcformat}


\title{Identification of criticality in neuronal avalanches: II. A theoretical and empirical investigation of the driven case}
 
%%%%%%%%%%%%%%%%%%%%%%%%%%%%%%%%%%%%%%%%%%%%%%
%%                                          %%
%% Enter the authors here                   %%
%%                                          %%
%% Ensure \and is entered between all but   %%
%% the last two authors. This will be       %%
%% replaced by a comma in the final article %%
%%                                          %%
%% Ensure there are no trailing spaces at   %% 
%% the ends of the lines                    %%     	
%%                                          %%
%%%%%%%%%%%%%%%%%%%%%%%%%%%%%%%%%%%%%%%%%%%%%%

\author{Caroline Hartley$^{1,2}$%
         \email{Caroline Hartley - caroline.hartley@ucl.ac.uk},
         Timothy J Taylor$^3$
         \email{Timothy J Taylor - tjt20@sussex.ac.uk},
        Istvan Z Kiss$^4$
         \email{Istvan Z Kiss - I.Z.Kiss@sussex.ac.uk},
         Simon F Farmer$^{5,6}$
         \email{Simon F Farmer - s.farmer@ucl.ac.uk}
        and
         Luc Berthouze\correspondingauthor$^{2,3}$
         \email{Luc Berthouze - L.Berthouze@sussex.ac.uk}
       }

%%%%%%%%%%%%%%%%%%%%%%%%%%%%%%%%%%%%%%%%%%%%%%
%%                                          %%
%% Enter the authors' addresses here        %%
%%                                          %%
%%%%%%%%%%%%%%%%%%%%%%%%%%%%%%%%%%%%%%%%%%%%%%

\address{%
    \iid(1) Centre for Mathematics and Physics in the Life Sciences and Experimental Biology, University College London, London, UK.\\
    \iid(2) Institute of Child Health, University College London, London, UK. \\
    \iid(3) Centre for Computational Neuroscience and Robotics, University of Sussex, Falmer, UK. \\
    \iid(4) School of Mathematical and Physical Sciences, Department of Mathematics, University of Sussex, Falmer, UK.\\
    \iid(5) National Hospital of Neurology and Neurosurgery, London, UK. \\
    \iid(6) Institute of Neurology, University College London, London, UK. 
}%

\maketitle

%%%%%%%%%%%%%%%%%%%%%%%%%%%%%%%%%%%%%%%%%%%%%%
%%                                          %%
%% The Abstract begins here                 %%
%%                                          %%  
%% Please refer to the Instructions for     %%
%% authors on http://www.biomedcentral.com  %%
%% and include the section headings         %%
%% accordingly for your article type.       %%   
%%                                          %%
%%%%%%%%%%%%%%%%%%%%%%%%%%%%%%%%%%%%%%%%%%%%%%

\begin{abstract}
        % Do not use inserted blank lines (ie \\) until main body of text.
The observation of apparent power-laws in neuronal systems has led to the suggestion that the brain is at, or close to, a critical state and may be a self-organised critical system. Within the framework of self-organised criticality a separation of timescales is thought to be crucial for the observation of power-law dynamics and computational models are often constructed with this property. However, this is not necessarily a characteristic of physiological neural networks - external input does not only occur when the network is at rest/a steady state. In this paper we study a simple neuronal network model driven by a continuous external input (i.e.\ the model does not have a separation of timescales) and analytically tuned to operate in the region of a critical state (it reaches the critical regime exactly in the absence of input - the case studied in the companion paper to this article). The system displays avalanche dynamics in the form of cascades of neuronal firing separated by periods of silence. We observe partial scale-free behaviour in the distribution of avalanche size for low levels of external input. We analytically derive the distributions of waiting times and investigate their temporal behaviour in relation to different levels of external input, showing that the system's dynamics can exhibit partial long-range temporal correlations. We further show that as the system approaches the critical state by two alternative `routes', different markers of criticality (partial scale-free behaviour and long-range temporal correlations) are displayed. This suggests that signatures of criticality exhibited by a particular system in close proximity to a critical state are dependent on the region in parameter space at which the system (currently) resides. 

\end{abstract}

\ifthenelse{\boolean{publ}}{\begin{multicols}{2}}{}

%%%%%%%%%%%%%%%%%%%%%%%%%%%%%%%%%%%%%%%%%%%%%%
%%                                          %%
%% The Main Body begins here                %%
%%                                          %%
%% Please refer to the instructions for     %%
%% authors on:                              %%
%% http://www.biomedcentral.com/info/authors%%
%% and include the section headings         %%
%% accordingly for your article type.       %% 
%%                                          %%
%% See the Results and Discussion section   %%
%% for details on how to create sub-sections%%
%%                                          %%
%% use \cite{...} to cite references        %%
%%  \cite{koon} and                         %%
%%  \cite{oreg,khar,zvai,xjon,schn,pond}    %%
%%  \nocite{smith,marg,hunn,advi,koha,mouse}%%
%%                                          %%
%%%%%%%%%%%%%%%%%%%%%%%%%%%%%%%%%%%%%%%%%%%%%%

%%%%%%%%%%%%%%%%
%% Background %%
%%

\section*{Introduction}

In recent years, apparent power-laws (i.e.\ where a power-law is the best model for the data using a model selection approach~\cite{clauset,klaus}) have been observed experimentally in neurophysiological data leading to the suggestion that the brain is a critical system~\cite{chialvo,lk01}. These observations have included that of neuronal avalanches - cascades of neuronal firing recorded \textit{in vivo} and \textit{in vitro} whose size and duration appear to follow power-law distributions~\cite{bp1,bp2,gireesh,hahn,petermann}. Recently it has been claimed that equivalent neuronal avalanche behaviour with the same power-law relationship can be identified in human MEG (magnetoencephalography) recordings~\cite{shriki}. On a wider scale, fluctuations in oscillation amplitude in human (adult and child) EEG (electroencephalography) and MEG  exhibit a power-law decay of the autocorrelation function of the signal - a property known as long-range temporal correlations (LRTCs)~\cite{lk01,lk04,nikulin04,nikulin05,smit,berthouze}. These observations and the idea that the brain is a critical system have drawn much attention as critical systems have been shown to exhibit optimal dynamic range and optimal information processing~\cite{kinouchi,shewreview}. Moreover, it has led to the hypothesis that brain dynamics may fit within the framework of self-organised criticality (SOC), i.e.\ a system that does not require external tuning of parameters to reach the critical state~\cite{bak,jensen,lk01}. 

While the observation of power-laws within neuronal activity may be attractive we must address the issue of whether (specifically) a neuronal system in the region of a critical state can produce this type of dynamics. Propagation of the spiking of neurons within a network has been interpreted within the context of percolation dynamics and the theory of branching processes~\cite{essam,harris}. A critical branching process is a process such that one active node will activate on average one other node at the next time-step and so one can discern how this would relate to neuronal systems whereby the system is critical if one active neuron on average activates one other neuron at the next time-step. A critical branching process will display power-law dynamics, however, a number of assumptions underlying branching processes do not hold true in neurophysiological systems. Firstly, the theoretical analysis of branching processes relies on full-sampling of the system. Full-sampling is unlikely to occur in the experimental setting and this can have a profound effect on the observed distributions~\cite{priesemann}. Additionally, re-entrant connections invalidate the standard theory of branching processes~\cite{harris} which brings into question the idea that neuronal systems can be modelled as critical branching processes. Moreover, the strict definition of a critical system is one that operates at a second order phase transition which applies only to systems with infinite degrees of freedom. Therefore, we may expect a critical system to exhibit an exact power-law distribution in the case of infinite size but what should we expect if the system is finite? As neuronal systems are necessarily finite this is an important question in the field but one that has yet to be fully addressed.  Within experimental results this fact has been accounted for by the concept of finite-size effects - where a power-law is observed up to a cut-off value~\cite{jensen,bp1,bp2,klaus}. This cut-off value has been suggested to coincide with the size of the system and distributions from networks of different sizes have been shown to exhibit an exact scaling relationship - a phenomenon known as finite-size scaling~\cite{klaus,bonachela}. However, the finite-size effect with a cut-off value at system size has been assumed without analytical derivation (though, see the companion paper to this article~\cite{taylor}, as described below) and the question of how a finite critical system behaves and the types of dynamics possible from such a system remains open in the field. Whether a finite-size system should display the same signatures of criticality as the system in the limit of system size is not known.

In the companion paper to this article~\cite{taylor} we examined a computational model of a finite neuronal system analytically tuned to its critical state, defined as a transcritical bifurcation. There we showed that the dynamics of the system, which by analogy with experimental neuronal avalanches could be termed avalanches (discrete cascades of neuronal firing), exhibited scaling which does not follow an exact power-law but does exhibit partial scale-free behaviour. We were able to show that the cut-off value is approximately the system size, as suggested experimentally by the finite-size effect, but is analytically related to the lead eigenvalue of the transition matrix (the matrix of all possible transitions at each simulation step). This is an important observation given that avalanches in systems with re-entrant connections could in principle be of infinite size and yet experimental observations have suggested that neuronal avalanches exhibit a finite-size cut-off~\cite{bp1,klaus}. Overall, the results suggested that finite systems at criticality exhibit signatures of critical systems dynamics but do not (at least in this instance) exhibit exact power-laws as had previously been suggested. 

While the system studied in the companion paper leads us to a greater understanding of the dynamics displayed by a finite neuronal system, there is still an important difference between the system studied there and physiological neuronal systems. In the companion paper the system was seeded by setting a single neuron in the network into the active state and an avalanche was defined as the firing that occurred until the network returned to a stable state (the fully quiescent state). After this point no more firing could occur until the system was reseeded. This imposed a strict separation of timescales, with all avalanches and neuronal firing occurring on a much faster timescale than the timescale of the `external input' reseeding the system. Many other computational models have also taken this approach~\cite{levina,bak,olami}, with a separation of timescales thought to be necessary for the observation of self-organised critical dynamics~\cite{bonachela}. While a separation of timescales is likely to occur in some natural systems such as earthquakes, where friction in the Earth's plates build up over the course of years but energy is released in a matter of minutes, this is not a physiologically realistic assumption for a neuronal system. External input (be it from the environment or other areas of the nervous system) will not arrive only once the neuronal population has returned to a set state. Before physiological recordings can be interpreted within the field of critical systems we must address the question of the types of dynamics that should be expected by not only a finite-size system but also a system that is driven by a physiologically realistic external input. Can a finite-size system without an explicit separation of timescales in the region of a critical regime exhibit markers of criticality? How might the external input to the system affect these markers?

Previous authors examining computational neuronal networks with continuous driving (i.e.\ no explicit separation of timescales) have observed power-law dynamics~\cite{kinouchi,ribeiro,rubinov,larremore}. In particular, Kinouchi and Copelli~\cite{kinouchi} and Larremore et al.~\cite{larremore} analytically determined the parameters required such that the model they studied was at criticality and displayed peak dynamic range, in fully connected networks and networks with a range of topologies, respectively. However, these authors did not explicitly examine the firing dynamics of the system in the region of the critical regime, concentrating on average activity levels. In a SOC system such as the sandpile model~\cite{bak} the waiting times (periods of inactivity between avalanches) have been shown to follow an exponential distribution~\cite{boffetta}. However, these waiting times are related to the reseeding of the system - sand is added to cells chosen at random and the next avalanche begins when a cell exceeds the threshold. In contrast, recent experimental work has shown that waiting times between neuronal avalanches in cultures have a distribution with two trends - a (short) initial power-law region thought to relate to neuronal up-states and a bump in the distribution at longer waiting times thought to relate to neuronal down-states~\cite{lombardi}. Could this difference in these waiting time distributions (between the SOC sandpile model and the neuronal avalanches in culture) be explained by the fact that physiological neuronal systems do not have a separation of timescales? 

As mentioned previously, another signature of criticality that has been reported in neural systems is the presence of LRTCs. In the majority of cases they have been observed in large scale neuronal signals such as human brain oscillations. Recent endeavours have been made to link these observations of scale-free behaviour on large scales with neuronal avalanches~\cite{poil,palva}. Poil and colleagues demonstrated in a computational neuronal network that power-law distributed avalanches and LRTCs in oscillations emerge concurrently. In addition, LRTCs have also been detected in the waiting times of bursts of activity in cultures~\cite{segev} and the discontinuous burst activity recorded from extremely preterm human neonates~\cite{hartley}. Thus, LRTCs have been demonstrated in discrete neuronal activity yet they have not been examined in the waiting times of neuronal avalanches themselves. While LRTCs in avalanche activity would not be possible in a seeded computational system (where the activity is initiated `by hand' and there is no memory within the system's dynamics) it is conceivable that a driven system, which is more akin to physiological networks which can display LRTCs, might display this type of dynamics in the waiting times of neuronal avalanches. 

In summary, in this paper we aim to address the following questions:
\begin{enumerate}
\item Assuming that the brain, or population of neurons under study, operates in the region of a critical regime can it be expected to display power-law statistics given that it is a finite-size system? If not what distribution should we expect? As discussed, this question was also addressed in the companion paper~\cite{taylor}, where we studied a system without an external input. However, here we specifically consider this question in the context of a driven system, i.e.\ with a non-zero external input and no explicit separation of timescales.
\item Can we expect a finite-size neuronal system in the region of a critical regime to exhibit other markers of criticality, and specifically the presence of LRTCs? Does the presence of LRTCs relate to that of power-law distributions? As described above, LRTCs have been observed in neurophysiological data sets. However, a systematic examination of how LRTCs may relate to other markers of criticality in neuronal systems is lacking. 
\item How are signatures of criticality (power-law distributions and LRTCs) affected by proximity to the critical regime? One might assume that a system which is closer to a critical regime may exhibit signatures of criticality, whereas a system that is further from the critical regime will not.  Importantly, our analysis shows that this assumption is in fact not (always) true.
\end{enumerate}

In this paper, as in the companion paper, we examine a purely excitatory stochastic neuronal model. As in the companion paper, a number of assumptions are made to simplify the model with the outcome that it is analytically tractable and therefore can be tuned to operate in the region of a critical regime. This approach is taken as it allows direct exploration of the above questions, which would not be possible with a more complex system. We begin by examining the distributions of avalanche size and duration, investigating the presence of scale-free behaviour. We also show that as the system approaches the theoretical critical regime by decreasing the external input, there is a change in the distributions of avalanche characteristics with the appearance of partial scale-free behaviour in avalanche size. It is important to note that the definition of avalanches strongly depends on the choice of binning method. In the literature different definitions of avalanches are used in models with seeded systems and with systems where the dynamics are continuous (including physiological recordings). We will return to this in the discussion.

Unlike in the companion paper where the system was seeded after each avalanche, the system studied here does not have an explicit separation of timescales. This allows us to additionally assess the waiting times, which are intrinsic to the system, and we are able to analytically derive the distribution of waiting times. We then investigate the presence of partial LRTCs in the empirically derived waiting times. Finally, we show that as the system size increases (and the system approaches the theoretical critical regime from a different route) the range over which the correlations extend also increases.  Overall we find that the system displays different signatures of criticality depending on the region of the parameter space around the critical regime.

\section*{The model}

As in the companion paper, we study a stochastic model based on that of Benayoun~et~al.~\cite{benayoun}. Though greatly simplified from a physiological neural network, the model is chosen as it is analytically tractable and thus enables direct derivation of the parameters such that there is a critical regime. With this approach it is therefore possible to assess the dynamics of a neuronal system in the region of (or at) a critical regime. While Benayoun~et~al. considered a network with both excitatory and inhibitory connections, we simplify the system further (as in the companion paper), considering a network with purely excitatory synaptic connections. As will be discussed later, this type of network can be set within the context of early brain development.

We consider a system of $N$ fully connected neurons, with each neuron in one of two states - active ($A$) or quiescent ($Q$). For a small time step $dt \rightarrow 0$ the probability of transition for a neuron between the two states is given by:
\[
P(Q\rightarrow A, \text{ in time }dt)=f(s_{i}(t))dt, \]
\[P(A\rightarrow Q, \text{ in time }dt)=\alpha dt,\]
where $s_{i}(t)=\sum_{j}\frac{w_{ij}}{N}a_{j}(t)+h_{i}(t)$ is the input to neuron $i$, $f$ is an activation function, $h_{i}(t)$ is the external input to neuron $i$, $w_{ij}$ is the connection strength from neuron $i$ to neuron $j$ and $a_{j}(t)=1$ if neuron $j$ is active at time $t$ and zero otherwise. Finally, $\alpha$ is a constant rate at which neurons change from the active to inactive (quiescent) state.

For analytical tractability and characterisation of the critical state we make the following additional simplifications:
\begin{enumerate}
\item The synaptic connection strengths are the same for all connections with $w_{ij}=w>0$.
\item The external input is constant to all neurons and at all simulation steps so that $h_{i}(t)=h>0$.
\item The activation function is linear with $f(x)=x$.
\end{enumerate}
While the first and third assumptions are the same as in the companion paper, we make the additional assumption of constant positive external input here as opposed to the companion paper where we examined the system with no external input ($h=0$). As the network is fully connected, and the system is closed so that $A+Q=N$ (where $A$ is the number of active neurons and $Q$ is the number of quiescent neurons), the system can be described by the mean field equation:
\[
\frac{dA}{dt}=\left( \frac{wA}{N}+h\right) \left( N-A \right) -\alpha A.
\]
As stated in the companion paper, we can use this equation to analyse the stability of the system about the fixed point and determine the parameters for which the system is at the threshold of stability, i.e.\ when the fixed point is critical. This threshold occurs when the eigenvalue ($\lambda$) of the fixed point is zero, which can alternatively be stated, borrowing terms from the epidemiology literature, as $R_{0}=1$ (the basic reproductive ratio). Moreover, this is also equivalent to a branching parameter of one. In the companion paper it was shown that with $h=0$, $R_{0}=\frac{w}{\alpha}$ and so for $R_{0}=\frac{w}{\alpha}=1 \Rightarrow \alpha=w$ the system is critical. 

Here we study the system in the presence of a positive external input, $h>0$. In this case the fixed point of the system is given by:
\[
-\frac{w}{N}A^{2}+wA+h(N-A)-\alpha A =0,
\]
and the eigenvalue at the fixed point is:
\[
\lambda =-2\frac{w}{N}A+w-h-\alpha.
\]
For a fixed point to be critical we require that both these equations be satisfied. However, solving them simultaneously we find that there are no real roots when $w,\ h,\ N>0$. This implies that there is no parameter region such that the system (with this activation function and positive external input) has a critical fixed point. However, considering again the case with no external input ($h=0$) for which the critical state occurred with parameters $\alpha =w$, if this system is driven by a `sufficiently low' level of external input it should still be within the region of the critical state. There has been some suggestion that the brain is not directly at a critical point but is in fact just very close to the critical regime and it has been speculated that the brain may actually be slightly supercritical~\cite{poil}.  Additionally, it been shown that a computational model of neuronal avalanches which follows a SOC approach~\cite{levina} is actually a system that `hovers' close to the critical state~\cite{bonachela}. Therefore, the question of how a finite driven system behaves in the region of a critical regime is pertinent to the neuroscience field. 

An additional motivation for considering a non-zero external input is the dynamic range ($\Delta$) of the system.  Larremore and colleagues~\cite{larremore} describe the dynamic range as ``the range of stimuli over which there is significant variation in the collective response of the network".  Kinouchi and Copelli~\cite{kinouchi} examined dynamic range in models of networks with uniform connectivity operating with \textit{discrete} time dynamics where multiple firings can occur within each time step.  They found that the dynamic range was maximised when the local branching ratio was equal to one.  Larremore and colleagues~\cite{larremore} considered a version of this model but with the introduction of heterogeneity in connections, showing that it is the lead eigenvalue, $\lambda$, of the connectivity matrix that governs the dynamic range and that the dynamic range is maximised when $\lambda = 1$.  In Appendix 1 we provide an analytic calculation for the dynamic range of our \textit{continuous} model.  Analogous to the results described above~\cite{kinouchi,larremore} the dynamic range is maximised when $R_0 = 1$ ($w/\alpha =1$ when $h=0$). This is illustrated in Fig.~\ref{fig:dynamicrange} where results from simulations are compared with the analytic solution.  It is important to emphasise that the parameterisation of the dynamic range is in terms of the value $R_0$ calculated for networks when there is no external input. When this parameterisation is such that $R_{0}=1 \Rightarrow \alpha =w$ (and therefore when the system is tuned to the critical state) external input to the system will give rise to dynamics for which the dynamic range is maximised. This point will be considered further in the discussion. 
\begin{figure}
\centering
        \includegraphics[width=0.8\textwidth]{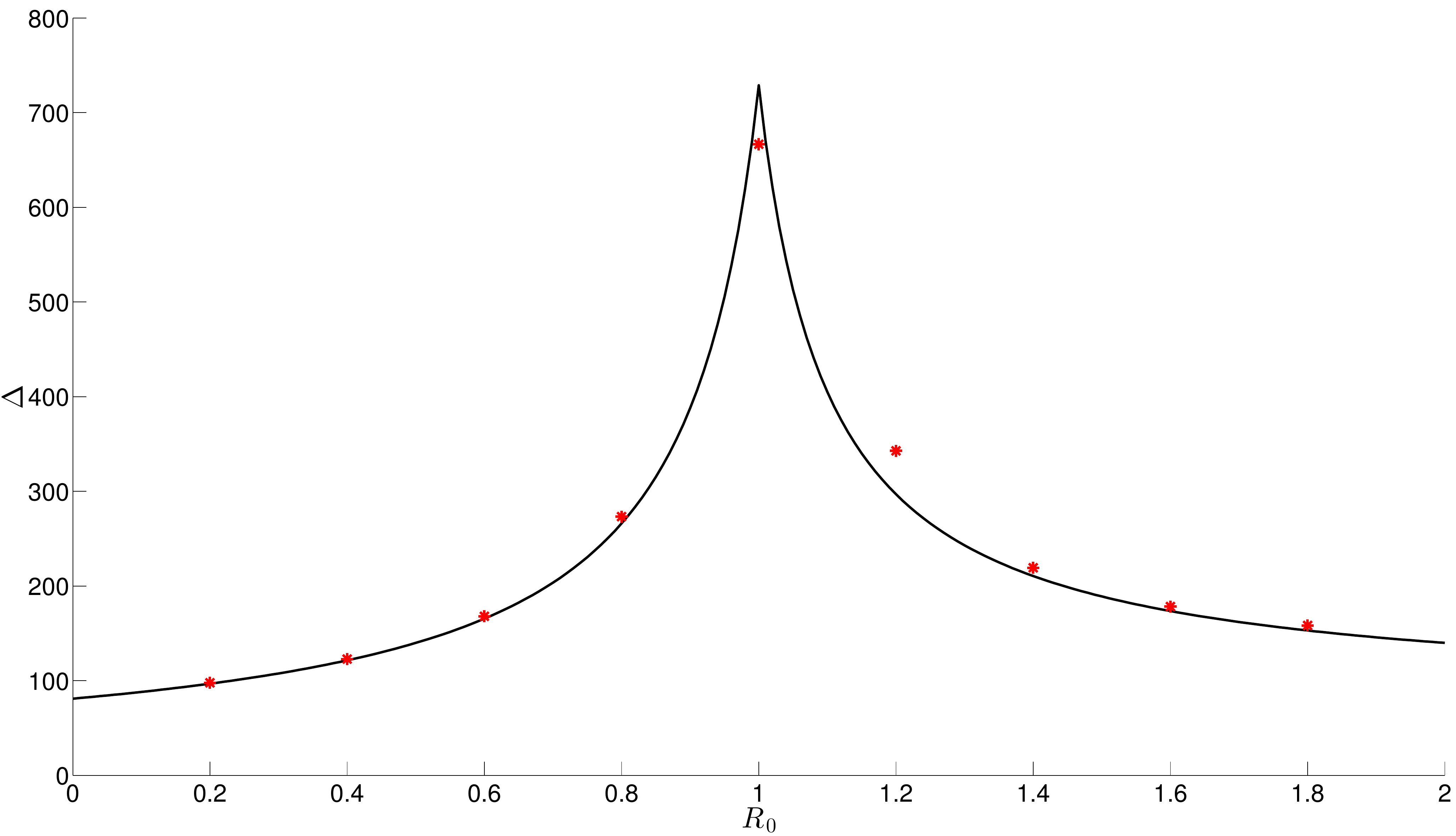}
    \caption{\textbf{$R_0$ versus $\Delta$.} Plot of $R_0$ versus $\Delta$ (the dynamic range - see Appendix 1) for both the analytic result (black line) and simulations (red dots).  For the comparable simulated result we average $10,000$ realisations that have run until time $t=200$.  To obtain a reasonable spread of $h$ we used the conjugation of the intervals $[0:0.002:0.2]$ and $[0.4:0.2:18]$.}
\label{fig:dynamicrange}
\end{figure}

Throughout this study we will examine the system in the presence of an external input of $h=1/N$ or less. This level of the external input is equivalent to setting a single neuron to the active state and so corresponds to seeding the system in the zero input case. We therefore deem this level of external input to be sufficiently low such that we could expect the system to remain within the region of the critical regime. As in the companion paper we set $w=\alpha =1$. With these parameters and with positive external input we find that the fixed point of the system is given by
\[
A=-\frac{hN}{2} \pm \sqrt{\frac{N^{2}h^{2}}{4}+N^{2}h},
\]
and the eigenvalue of this fixed point is given by
\[
\lambda=-\sqrt{h^{2}+4h}.
\]
With lower levels of external input the system approaches the critical regime (see Fig.~\ref{fig:eigenvalueNh}). Note that this approach is in fact from a slightly subcritical state given these values of $\alpha$ and $w$ and with positive external input. Under these conditions it is not possible to consider an approach from a supercritical regime with a positive eigenvalue. 

As described above, we (initially) set $h=1/N$. With this level of external input:
\[
A=-\frac{1}{2} \pm \sqrt{\frac{1}{4}+N},
\]
and the eigenvalue of the fixed point is given by
\[
\lambda=-\sqrt{\frac{1}{N^{2}}+\frac{4}{N}}.
\]
As $N \rightarrow \infty$, $\lambda \rightarrow 0$ (see Fig.~\ref{fig:eigenvalueNh}). Thus, for this level of the external input ($h=1/N$), as the system size ($N$) increases the system approaches the critical state (as the system reaches the critical state exactly when the eigenvalue $\lambda =0$). We will examine the effect on the dynamics of decreasing the external input, thereby allowing the system to approach the critical regime. We will also investigate an alternative route to the critical regime by increasing system size at constant (overall) level of external input.

\begin{figure}
\centering
	\includegraphics[width=150mm]{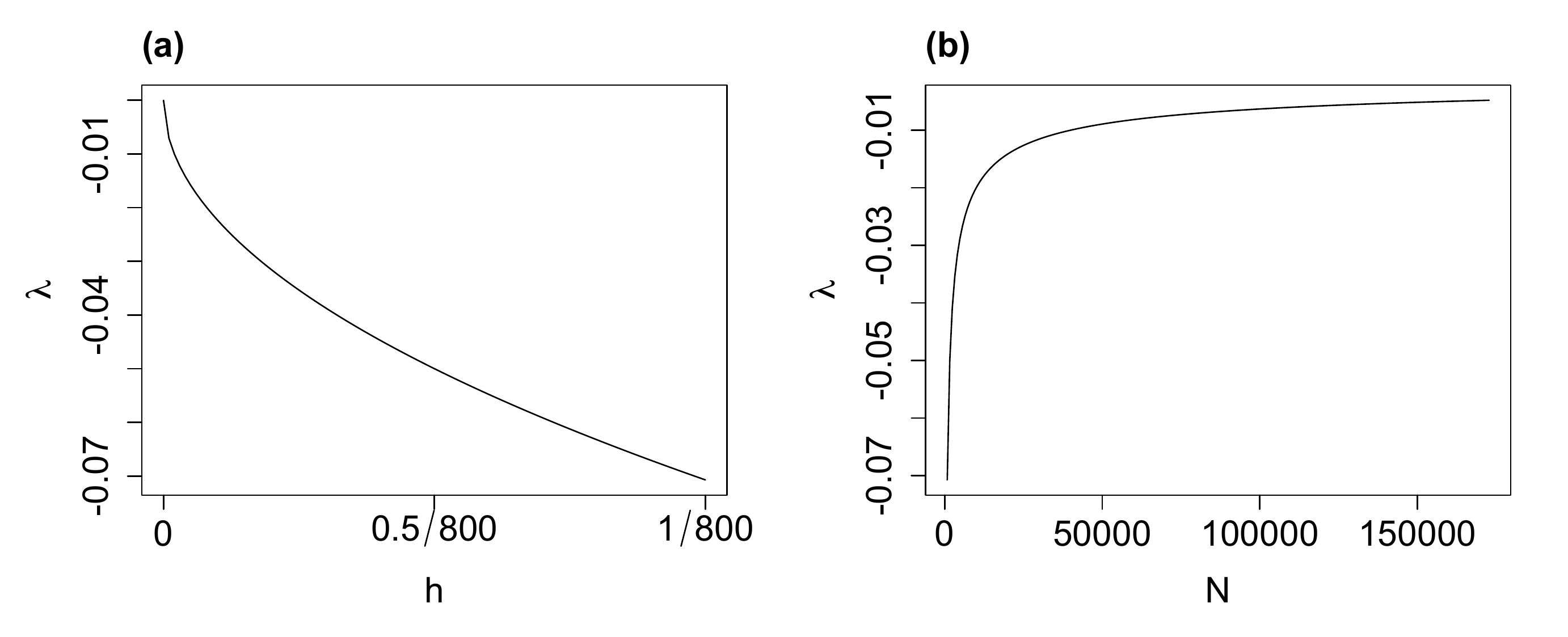}
\caption{\textbf{The eigenvalue of the system compared with the level of external input and system size.} (a) With $w=\alpha$ the eigenvalue decreases with lower levels of external input $h$, with the system reaching the critical regime in the absence of external input ($\lambda=0$ i.e.\ $R_{0}=1$, the case studied in the companion paper). (b) With $h=1/N$ and $w=\alpha$ the eigenvalue of the fixed point $\lambda \rightarrow 0$ as $N \rightarrow \infty$. Thus, the system approaches the critical state as the system size increases.}
\label{fig:eigenvalueNh}
\end{figure}

\subsection*{Model simulations and burst analysis}

As in the companion paper and in Benayoun et al.~\cite{benayoun}, simulations of the network dynamics were carried out using the Gillespie algorithm for stochastic simulations~\cite{gillespie}. Briefly, at each step in the simulation
\begin{itemize}
\item The total transition rate $r$ for all the neurons within the network is calculated, with $r=r_{aq}+r_{qa}$ where $r_{aq}$ is the total rate of active $\rightarrow$ quiescent transitions and is given by $r_{aq}=\alpha A$ and $r_{qa}$ is the total rate of all quiescent $\rightarrow$ active transitions which is given by $r_{qa}=f(s_{i})(N-A)$. 
\item The time to the next transition $dt$ is selected at random from an exponential distribution of rate $r$.
\item The type of transition is selected by generating a random number $n \in [0,1]$. If $n<\frac{r_{aq}}{r}$ then a randomly chosen active neuron becomes quiescent, otherwise a (randomly chosen) quiescent neuron switches to the active state.
\end{itemize}
At each step in the simulation a single neuron makes a transition, though the rate at which transitions occur changes and so the simulation step changes. If the network is in a fully quiescent state ($Q=N$) then, with positive external input, $r_{aq}=0$ but $r_{qa}=hN$ and consequently there will necessarily be a transition of a neuron from the quiescent to the active state. Similarly, when the network is in the fully active state ($A=N$) $r_{qa}=0$ but $r_{aq}=\alpha N$ and so there will necessarily be a transition of a randomly chosen neuron from the active to the quiescent state. From all other starting points transitions from the active to the quiescent or from the quiescent to the active state are possible. Thus, from all network states one neuron will change state. This is unlike the companion paper where with no external input the network must be seeded when in the fully quiescent state. Instead in this case network dynamics are continuous (i.e.\ no re-seeding is required) and are of finite length only in-so-far as they are restricted by simulation lengths. 

We define a neuron as firing at the first time step at which the neuron switches from the quiescent to the active state. Fig.~\ref{fig:examplesims} shows raster plots of network firings for the first 1 second of simulations with three different levels of the external input. As was described above, unlike in the companion paper where there was no external input, the dynamics continue even if the system reaches the fully quiescent state. Interestingly, we can also notice that the network dynamics appear to exhibit burst-like behaviour, with periods of high neuronal firing interspersed with periods without network firing. It is important to realise that these bursts are intrinsic to the system and are not directly related to the dynamics of the external input (the input is constant to all neurons in each of the simulations) nor due to a saturation of the network - the bursts themselves consist of different numbers of neuronal firing. In all three cases the parameters are set to the critical state (with no external input). With lower levels of the external input the system approaches the critical regime and we see that the bursts become further apart and more distinct. We will characterise these dynamics below. (See also Appendix 2 where we examine driving the system from subcritical and supercritical states.)

\begin{figure}
\centering
	\includegraphics[width=150mm]{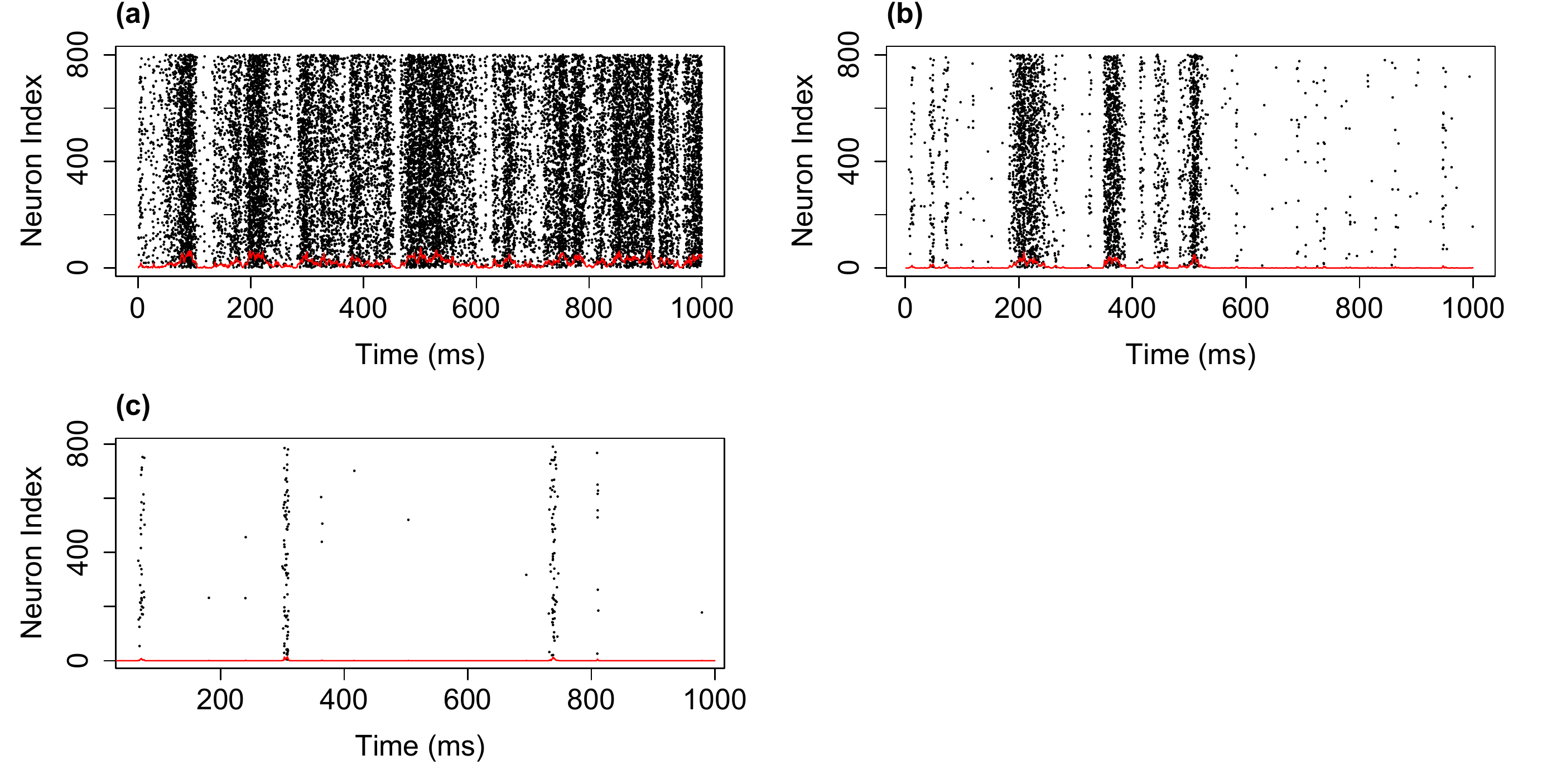}
\caption{\textbf{Raster plots of the network dynamics for different levels of the external input.} Neuronal firing across 1 second of a simulation with an external input of (a) $h=1/N$, (b) $h=0.1/N$ and (c) $h=0.01/N$. For all three simulations $N=800, \ \alpha =w=1$ and the red line indicates the rate of firing in 1~ms bins. As can be observed, as the level of the external input decreases the firing rate decreases and the time between avalanches increases.}
\label{fig:examplesims}
\end{figure}

These burst dynamics are analogous to the neuronal avalanches observed experimentally in that they are discrete cascades of firing. Neuronal avalanches observed experimentally in physiological networks are so called because they have sizes which are distributed according to a power-law and while the size distribution of the burst activity in this network has yet to be presented we will refer to the activity throughout the rest of this paper as avalanches due to their discrete burst behaviour. To determine the distribution of the avalanches we divided the activity into individual avalanches using the approach of Benayoun et al.~\cite{benayoun}. This method divides consecutive neuronal spiking between any two neurons within the network into separate avalanches if the time difference between the spikes is greater than the average difference ($\delta t$) between consecutive spikes within the simulation. This approach (referred to later in the text as the binning method) is similar to the method used to define neuronal avalanches within physiological data~\cite{bp1,bp2} - though the choice of binning method will be discussed later in the paper. It is important to note that no binning approach was needed in the companion paper since an avalanche was naturally defined as all firing that occurred before the network reached the fully quiescent state and was reseeded. This has been used as a standard classification for discontinuous data, stemming from the sandpile model of criticality~\cite{bak}. However, as the firing dynamics here continue for the entire simulation it was instead appropriate to use an approach that had been used previously for continuous dynamics. 

Throughout the remainder of this paper we examine characteristics of these avalanches: namely the size and duration of avalanches as well as the inter-avalanche intervals (IAIs). The size of an avalanche is defined (in the standard way) as the number of firings within the avalanche. If a single neuron fires more than once within a single avalanche it is also counted more than once. The duration of an avalanche is defined as the time between the start of the avalanche (the first neuron firing) and the end of the avalanche. Note that if the avalanche consists of a single neuron firing then the duration of the avalanche is 0 (and the size of the avalanche is 1). Similarly, an IAI is defined as the time between the end of one avalanche and the start of the next avalanche, i.e.\ the waiting time between avalanches. Note that the minimum IAI is bounded below by $\delta t$ as a separation between two consecutive spikes of greater than $\delta t$ defines separate avalanches. 

\subsection*{Distributions of avalanche size and duration}

Fig.~\ref{fig:distsize} shows the distributions of avalanche size and duration from example simulations for the three different levels of external input investigated -- this can be compared with Fig. 3 of the companion paper~\cite{taylor} which shows the avalanche size distribution in the absence of external input. With lower levels of external input the system approaches the critical regime and the distributions of avalanche size appear scale-free across a range of scales. The distribution approaches the distribution found in the companion paper for the system exactly tuned to the critical state. With $h=0.01/N$ (i.e.\ the lowest level of eternal input) the exponent of the fitted power-law is approximately 1.5, see Fig.~\ref{fig:distsize}(c), which is consistent with experimentally observed neuronal avalanche sizes~\cite{bp1,bp2}. However, for higher levels of the external input this scale-freeness of the distribution is lost which coincides with moving away from the critical regime. In the case of avalanche duration a similar relationship with the critical regime is seen with a scale-free portion in the middle ranges of the distribution (between approximately 2 and 50 ms) with lower levels of external input. Thus, for lower levels of external input, when the system approaches the critical regime, the distributions, in particular that of avalanche size, exhibit partial scale-free behaviour. 

\begin{figure}
\centering
	\includegraphics[width=150mm]{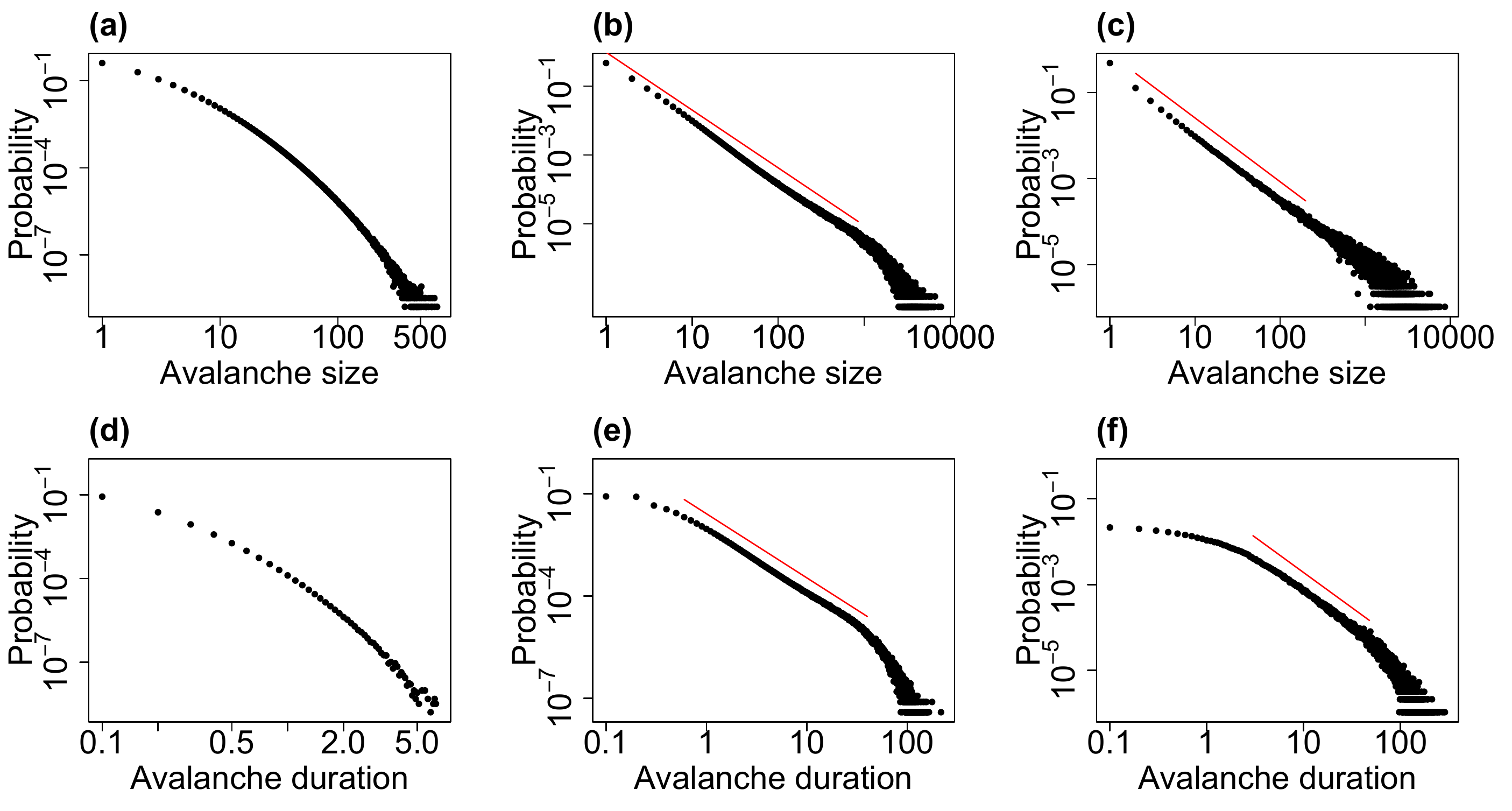}
\caption{\textbf{Distributions of avalanche size and duration for varying levels of the external input.} The distributions of avalanche size (a,b,c) and duration (d,e,f) from simulations with $h=1/N$ (a,d), $h=0.1/N$ (b,e) and $h=0.01/N$ (c,f). As the level of the external input is decreased the system approaches the critical regime. For all simulations $N=800,\ \alpha =w=1$ and the distributions are pooled from 10 simulations each of length $10^{4}$ seconds. The red lines indicates linear fits on the double logarithmic scale (where appropriate), i.e.\ fitted power-laws with exponents of (b) 1.68, (c) 1.48, (e) 1.88 and (f) 1.64.}
\label{fig:distsize}
\end{figure}

It is worth considering what leads to the changes seen in the distributions as the level of external input is varied. As stated, as the level of external input decreases, the system approaches the critical regime and so it is perhaps not surprising that signatures of criticality (i.e.\ scale-free behaviour) emerge in the distribution of avalanche size. Examining the raster plots of firing for the different levels of external input, see Fig.~\ref{fig:examplesims}, we see that at lower levels the avalanches are further apart and more distinct. While the external input itself is continuous, at lower levels of external input there is a separation of timescales, where one avalanche always finishes well before the next avalanche begins. The distribution therefore appears to follow similar characteristics to a system with a built-in separation of timescales and we confirm that the distribution is similar to that found in the companion paper (in which the model had an explicit separation of timescales) where an exponent close to 1.5 was also observed for the distribution of avalanche size. As the external input is increased there are no longer such distinct periods between avalanches. This leads to a superposition effect, with the next (actual) avalanche starting before the previous avalanche has finished (i.e.\ a new network cascade is initiated before the previous one has finished). This leads to these `avalanches' being defined using the binning approach as a single avalanche (see discussion). The scale-free behaviour in the distributions of avalanche size and duration is therefore lost.

\subsection*{Theoretical derivation of the distribution of the IAIs and comparison with simulated data}

The temporal patterning of activity within networks of neurons has long been investigated as a property of key importance, with rate and temporal coding suggested as potential substrates for information propagation. As it remains to be fully determined how different neuronal firing properties may lead to information transfer we suggest that in addition to the distribution of avalanches sizes the intervals between avalanches need to be considered as a functional entity in their own right. In this section we derive the theoretical distribution of IAIs and compare it with results from simulations.

We begin by noting that a single IAI is a period during which there is no neuronal firing, i.e.\ neurons can only be switching from the active to the quiescent states or an IAI may be a period with a single quiescent to active transition which is preceded by another quiescent to active transition. Let us initially ignore the fact that there is a minimum duration ($\delta t$) of an IAI and first consider the distribution of all consecutive active to quiescent transitions (we will return to the distribution of single quiescent to active transitions later). 

\subsubsection*{Distribution of consecutive active to quiescent transitions}
Let $N_{0}$ be the number of \textit{active} neurons at a time point in the simulation. After a single simulation step the number of active neurons will be $N_{0}+1$ or $N_{0}-1$, as at every simulation step only one neuron makes a transition. Let $q_{i}$ be the probability that an active neuron goes back to the quiescent state given that there are $i$ active neurons. Note that from the transition rates:
\[
q_{i}=\frac{\alpha i}{(\frac{w}{N}i+h)(N-i)+\alpha i}=\frac{\alpha iN}{(wi+hN)(N-i)+\alpha iN}.
\]
Starting with $N_{0}$ active neurons the probability that there are $N_{0}-1$ active neurons after a single simulation step is $q_{N_{0}}$ and the probability that there are $N_{0}+1$ active neurons is $1-q_{N_{0}}$. Given these probabilities we can construct a probability tree, shown in Fig.~\ref{fig:probtree}, particularly concentrating on the portion of the tree corresponding to active to quiescent transitions, i.e.\ those transitions that form a period of consecutive active to quiescent transitions. (Note that this probability tree focuses on different aspects of the model to that of the probability tree in the companion paper.)

\begin{figure}[htp]
\centering
\includegraphics[width=60mm]{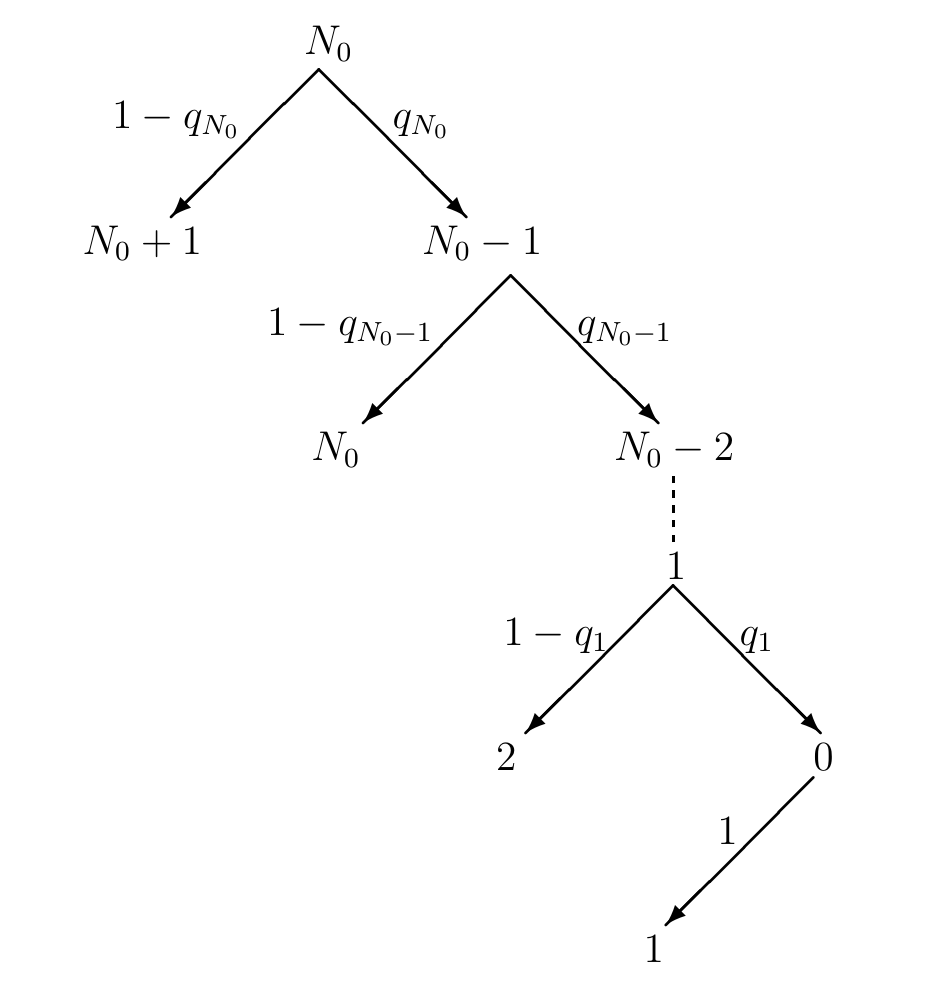}
\caption{\textbf{Probability tree of consecutive active to quiescent transitions.} Starting from a state with $N_{0}$ active neurons the probability tree diagram indicates the possible transitions from each state specifically concentrating on the active to quiescent transitions. The probability $q_{i}$ of each transition is as indicated in the main text and is dependent on the number of active neurons, $i$.}
\label{fig:probtree}
\end{figure}

From this tree approach we can calculate that the probability of \textit{exactly} $k$ consecutive active to quiescent transitions (note that to consist of \textit{exactly} $k$ active to quiescent transitions the transition sequence must be ended by a quiescent to active transition):
\begin{align}
P(IAI_{k})=p(N_{0},k)=(1-q_{N_{0}-k})\prod_{j=0}^{k-1}q_{N_{0}-j}.
\label{eq:probk}
\end{align}

The duration of this period of consecutive active to quiescent transitions is given by the sum of the times for each of these $k$ transitions (plus the time for the quiescent to active transition). As the Gillespie algorithm is used for simulations, at each simulation step the time to the next transition is drawn randomly from an exponential distribution with rate $r$ (see above), where $r$ is dependent on the number of active neurons and so changes at each simulation step. The duration of consecutive active to quiescent transitions is therefore the sum of exponentially distributed variables drawn from distributions of different rates, i.e.\ the distribution of consecutive active to quiescent transitions is a hypoexponential. Thus, the duration distribution, $f(x,N_{0},k)$, of consecutive active to quiescent transitions of length $x$, consisting of $k$ transitions, ending with an additional quiescent to active transition and starting from $N_{0}$ active neurons is~\cite{ross}:
\begin{align}
f(x,N_{0},k)=\sum_{j=0}^{k}r_{N_{0}-j}e^{-r_{N_{0}-j}x}\left( \prod_{i=0,i\neq j}^{k}\frac{r_{N_{0}-i}}{r_{N_{0}-i}-r_{N_{0}-j}} \right),
\label{eq:hypoexp}
\end{align}
when $r_{N_{0}-i} \neq r_{N_{0}-j}$  and where $r_{m}$ is the total transition rate for all neurons within a network with $m$ active neurons and is the rate of the exponential distribution from which the time to the next transition is randomly drawn. This equation holds provided that $r_{N_{0}-i} \neq r_{N_{0}-j}\ \forall i,j$. If this is not the case and there exists $A,B\in \{1, \hdots N\}$ such that $(\alpha + w -h)/w = A+B \ \Rightarrow \ r_A = r_B$ then we use the more general form -- assuming there are $a$ distinct rates, which we label $\beta_{1},\ \beta_{2}, \ldots, \beta_{a}$ that occur $c_{1},c_{2}, \hdots , c_{a}$ times respectively (i.e.\ $c_{1}+c_{2}+ \hdots c_{a}=k+1$) -- given by~\cite{scheuer}:
\begin{align}
f(x,N_0,k) &= B \sum_{k=1}^{a}{ \sum_{l=1}^{c_k} \frac{\phi_{k,l}(-\beta_k)x^{c_k-l}e^{-\beta_k x}}{(c_k-l)!(l-1)!}},
\end{align}
where
\begin{align*}
B &= \prod_{j=1}^a \beta_{j}^{c_j} \text{ and }  \phi_{k,l}(t) = \frac{d^{t-1}}{dt^{t-1}}\prod_{j=1,j \neq k}^{a}{(\beta_j + t)^{-c_j}}.
\end{align*}
Whilst this involves higher order derivatives a closed-form solution is provided by Amari and Misra~\cite{amari}. 

From equation~\ref{eq:probk} we know the probability of $k$ consecutive active to quiescent transitions. This equation holds true for any $k$ up to $k=N_{0}$, which is the maximum number of consecutive active to quiescent transitions as the fully quiescent state is then reached. Therefore, the distribution, $F(x,N_{0})$, of consecutive active to quiescent transitions of duration $x$ starting with $N_{0}$ active neurons but consisting of any number of transitions is a weighted sum of hypoexponentials:
\begin{align}
F(x,N_{0})=\sum_{k=1}^{N_{0}}f(x,N_{0},k)p(N_{0},k).
\label{eq:hypo}
\end{align}

\subsubsection*{Probability distribution of the initial number of active neurons}
Finally, to calculate the full probability distribution of consecutive active to quiescent transitions for a network of set system size, $N$, we must combine equation~\ref{eq:hypo} with the probability of the initial number of active neurons being equal to $N_{0} \in \{ 1,2,... N\}$ (note that $N_{0}=0$ is not considered as the next transition will necessarily be an activation). To determine this probability, let us first consider the simple case of $N=3$. We assume that the simulation starts from a state with no active neurons. Fig.~\ref{fig:n3tree} shows all possible transitions between the number of active neurons in a network of this size.

\begin{figure}[htp]
\centering
\includegraphics[width=50mm]{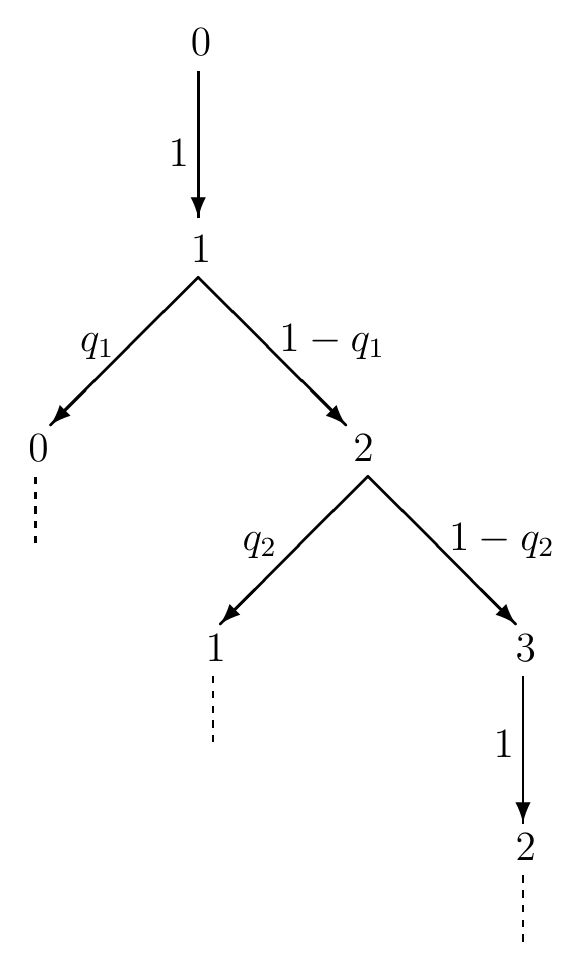}
\caption{\textbf{Probability tree of all possible transitions in a network of size $\mathbf{N=3}$.} Simulations start from a state with no active neurons: $N_{0}=0$. The diagram shows the possible transitions at each step, along with the probability of making that transition. The probabilities are as defined in the main text with $q_{i}$ being the probability of a neuron switching from the active to the quiescent state given $i$ initially active neurons. Dotted lines indicate transitions that are already shown elsewhere in the tree and so the tree shown here completely describes all possible transitions in a network of this size.}
\label{fig:n3tree}
\end{figure}

From this probability tree the probabilities, $P(i)$, of the number of active neurons being equal to $i$, where $i \in \{ 0,1,2,3 \}$ are given by:
\begin{align}
P(0)= &\ q_{1}P(1), \notag \\
P(1)= &\ P(0)+q_{2}P(2), \\ 
P(2)= &\ (1-q_{1})P(1)+P(3), \notag \\
P(3)= &\ (1-q_{2})P(2), \notag
\label{eq:standardeqnsN3}
\end{align}
(assuming a steady state has been reached such that the probabilities are time independent). Rearranging and substituting to write the equations in terms of $P(1)$:
\begin{align}
P(0)= &\ q_{1}P(1), \notag \\
P(2)= &\ \frac{(1-q_{1})}{q_{2}}P(1), \\
P(3)= &\ \frac{(1-q_{1})(1-q_{2})}{q_2}P(1). \notag
\label{eq:eqnsN3}
\end{align}
Furthermore, the sum of all the probabilities must equal 1 and so
\begin{align}
\left( q_{1}+1+\frac{(1-q_{1})}{q_{2}}+\frac{(1-q_{1})(1-q_{2})}{q_2} \right) P(1) =1.
\end{align}
Therefore,
\begin{align}
P(1)=\frac{q_{2}}{q_{1}q_{2}+q_{2}+(1-q_{1})+(1-q_{1})(1-q_{2})}.
\end{align}
By substituting this value back into the set of equations 6 the probabilities for the full system can be calculated.

\subsubsection*{Generalisation to a system of any size $N$}
From considering this simple example we can generalise to derive the probabilities of the number of active neurons for a system of any size $N$. Firstly, as in equation 4 the probabilities can be written as (again assuming a steady state):
\begin{align}
P(0)= & \ q_{1}P(1), \notag \\
P(1)= &\ P(0)+q_{2}P(2), \notag \\ 
P(2)= &\ (1-q_{1})P(1)+q_{3}P(3), \notag \\
\vdots \notag \\
P(k) = &\ (1-q_{k-1})P(k-1)+q_{k+1}P(k+1),  \label{eq:probsN} \\
\vdots  \notag \\
P(N-1)= &\ (1-q_{N-2})P(N-2) + q_{N}P(N), \notag \\
P(N) = &\ (1-q_{N-1})P(N-1), \notag 
\end{align}
where $q_{N}=1$ but it will remain in the equations so as to aid notation.
Rearranging gives
\[
P(2)=\frac{(1-q_{1})}{q_{2}}P(1), 
\]
and by induction:
\begin{align}
P(k+1) = &\ \frac{1}{q_{k+1}}\left( P(k)-(1-q_{k-1})P(k-1) \right) \notag \\
	= &\ \frac{(1-q_{1})(1-q_{2}) \hdots (1-q_{k})}{q_{2}q_{3}\hdots q_{k+1}} P(1). 
\label{eq:probsNk}
\end{align}
Summing all the probabilities and setting this equal to 1:
\begin{align}
P(1) = \frac{q_{2}q_{3} \hdots q_{N}}{q_{1}q_{2} \hdots q_{N} +q_{2}q_{3} \hdots q_{N} +(1-q_{1})q_{3}\hdots q_{N} + \hdots +(1-q_{1})(1-q_{2})\hdots (1-q_{N})}.
\label{eq:p1N}
\end{align}

Having determined these probabilities we then need to take into account the fact that the consecutive active to quiescent transitions must be preceded by a quiescent to active transition, i.e.\ they must be preceded by a neuron firing (otherwise they would be a chain of $k+1$ consecutive active to quiescent transitions and so included elsewhere in the distribution). We are therefore only interested in the probability $P_{A}(N_{0})$ of the number of active neurons being equal to $N_{0}$ given that a quiescent to active transition has just occurred. Considering again the probability tree, Fig~\ref{fig:n3tree}, we find that these probabilities, are given by:
\begin{align}
P_{A}(0)= &\ 0, \notag \\
P_{A}(1)= &\ P(0), \notag \\
\vdots \notag \\
P_{A}(k)= &\ (1-q_{k-1})P(k-1), \\
\vdots \notag \\
P_{A}(N)= &\ (1-q_{N-1})P(N-1), \notag 
\end{align}
where we make use of the previously defined probabilities $P(k)$. From these probabilities the full probability distribution of the duration of consecutive active to quiescent transitions can be calculated. As was shown above, for a set initial number of active neurons $N_{0}$, the probability distribution of consecutive active to quiescent transitions is given by a weighted sum of hypoexponentials, see equation~\ref{eq:hypo}. This can then be  further weighted by the probability $P_{A}(N_{0})$ that the initial (at the start of the sequence of transitions) number of active neurons is equal to $N_{0}$ and the previous transition was quiescent to active. Thus, the overall probability distribution of consecutive active to quiescent transitions is given by:
\begin{align}
\wp (x) = \sum_{i=0}^{N}\left( P_{A}(i) \sum_{m=1}^{i}f(x,i,m)p(i,m) \right).
\label{eq:fullprobdist}
\end{align}

To confirm that this theoretically derived distribution compares with results from simulations, we determined the distribution of the lengths of periods of any consecutive active to quiescent transitions from simulations. Fig.~\ref{fig:theorsim} shows the good agreement between the distribution of consecutive active to quiescent transitions from a simulation with $N=50$ and the theoretical distribution. 

\begin{figure}
\centering
	\includegraphics[width=120mm]{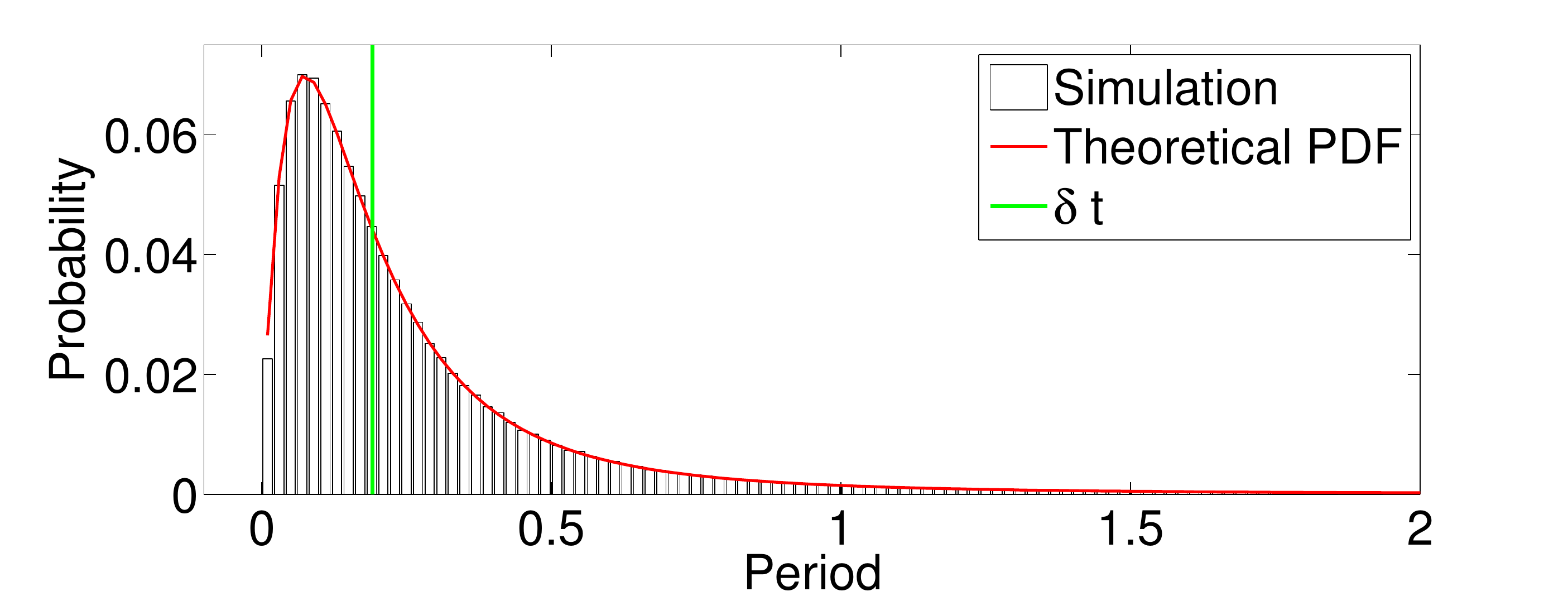}
\caption{\textbf{Theoretical and simulated distributions of periods of consecutive active to quiescent transitions.} The simulated distribution (black) is compared with the theoretically derived probability density function (shown in red, see equation~\ref{eq:fullprobdist}). For both distributions $\alpha =1,\ w=1,\ N=50$ and $h=1/N$. The mean difference between consecutive spikes ($\delta t$) within the simulation (green) is used to define avalanches through the binning approach described in the main text. Thus, the portion of the distribution for which the length of active to quiescent transitions are greater than this average time between consecutive spikes form the distribution of IAIs when combined with the distribution of single quiescent to active transitions.}
\label{fig:theorsim}
\end{figure}

\subsubsection*{Distribution of single quiescent to active transitions}
As was described above, a single period in between two neurons firing can also be an IAI, provided that the duration of this IAI is longer than the average time between spikes as accounted for below. Note, however, that only single periods are considered, as consecutive periods necessarily include neurons switching to the active state and therefore cannot form part of an IAI. Thus, the distribution of IAIs should also take into account the distribution of single quiescent to active transitions. As we make use of the Gillespie algorithm, the duration distribution of these single transitions is an exponential with rate given by the total transition rate, which is dependent on the number of active neurons, $N_{0}$. This is then weighted by the probability of a quiescent to active transition given $N_{0}$ active neurons (i.e.\ by $1-q_{N_{0}}$) and additionally weighted by the probability of starting with $N_{0}$ active neurons following a quiescent to active transition as calculated above. Thus, the probability distribution of single quiescent to active transitions of length $x$ is given by:
\begin{align}
\rho=\sum_{i=0}^{N}\left( P_{A}(i) (1-q_{i}) r_{i}e^{-r_{i}x} \right).
\end{align}
Fig.~\ref{fig:singleqadist}(a) shows the good agreement between simulated distribution of single quiescent to active transitions and theoretical distribution. 

\subsubsection*{The IAI distribution}
As discussed above, the IAI distribution combines these two distributions - the distribution of consecutive active to quiescent transitions and the distribution of single quiescent to active transitions. This combined distribution, along with simulated values, is shown in Fig.~\ref{fig:singleqadist}(b). As was described above, avalanches are defined from the network firing pattern as consecutive spikes where the time difference between them is no greater than the average time difference between consecutive spikes, $\delta t$, within the network. Thus, the minimum IAI is bounded below by $\delta t$ and all consecutive active to quiescent transitions or single quiescent to active transitions whose total duration is greater than $\delta t$ will be an IAI. Thresholding the combined distribution at $\delta t$ determines the IAI distribution. Fig.~\ref{fig:IAIdists}(a) shows theoretical and simulated IAI distributions displayed on a double logarithmic scale. Despite the fact that the distribution is not a power-law (theoretically we know that it is a weighted sum of hypoexponentials), it appears scale-free over a range of scales on this double logarithmic scaling. As we will show below, the distribution can also pass statistical tests for power-law distributions, suggesting partial scale-free behaviour of the system close to the critical regime.

\begin{figure}
\centering
	\includegraphics[width=130mm]{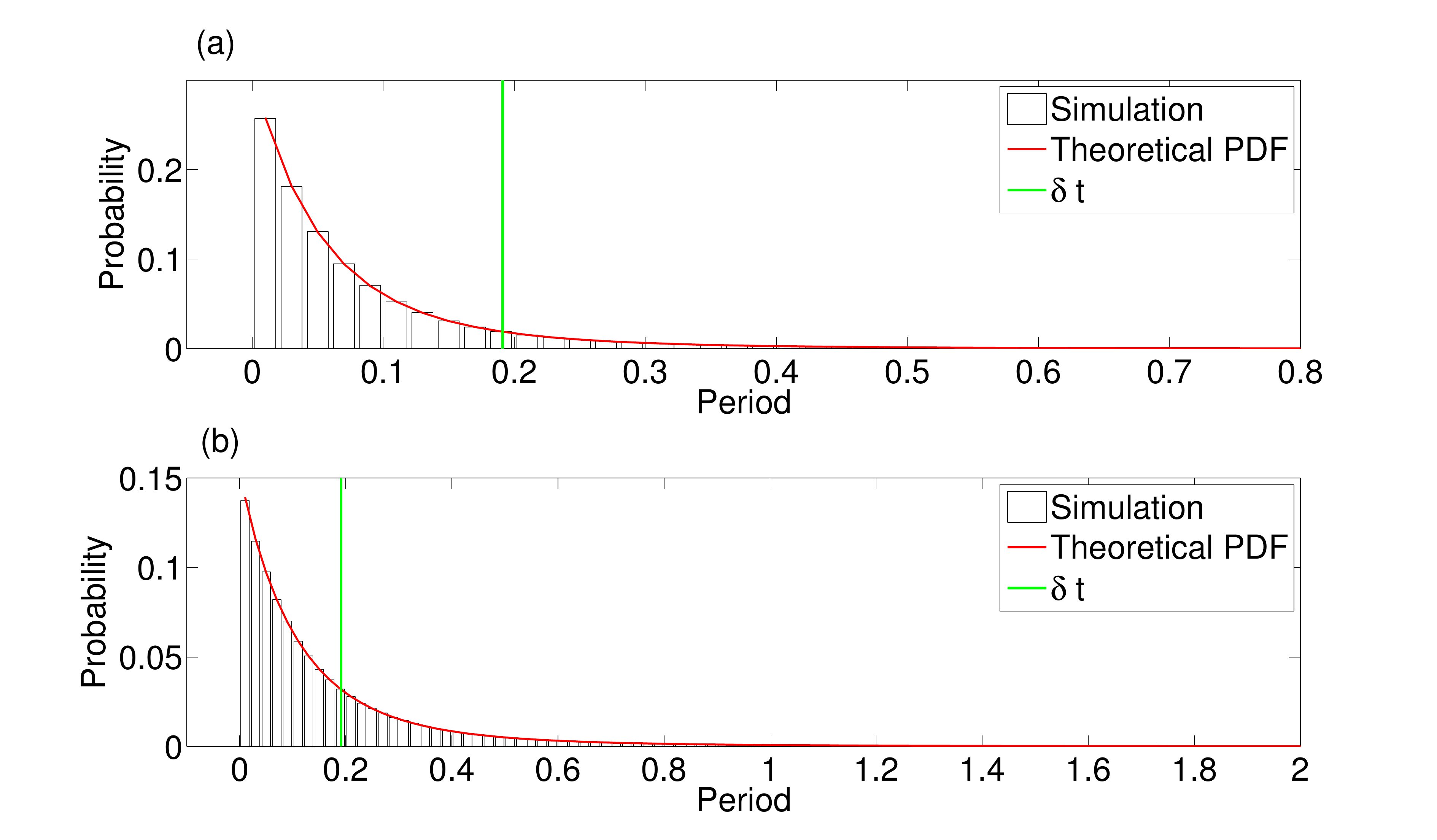}
\caption{\textbf{Theoretical distributions of single quiescent to active transitions and the combined distributions.} Theoretical (red) and simulated (black) probability distributions for (a) single quiescent to active transitions and (b) this distribution combined with the distribution of consecutive active to quiescent transitions (see Fig.~\ref{fig:theorsim}). In both cases $\alpha =1,\ w=1,\ N=50$ and $h=1/N$. The green line indicates the average time between consecutive spikes ($\delta t$) within the simulations. Thresholding the combined distribution (b) at this level determines the IAI distribution.}
\label{fig:singleqadist}
\end{figure}

Fig.~\ref{fig:IAIdists} also shows the theoretical and simulated distributions for lower levels of external input. With lower levels of external input (as the system approaches the critical regime) the average IAI increases and the distribution changes, no longer exhibiting scale-free behaviour. Even when considering the same scale for all levels of the external input (IAIs in the region of 0.05-5 ms) it is only for $h=1/N$ that the distribution is scale-free. Indeed, at the lowest level of input $h=0.01/N$ the distribution is in fact best fit by an exponential, in this case $y=0.028e^{-0.01x}$ as seen in Fig.~\ref{fig:IAIexponential}, indicating the loss of the scale-free behaviour in the distribution. Thus, scale-free behaviour in the case of the IAI distribution does not increase with proximity to the critical regime. 

\begin{figure}
\centering
	\includegraphics[width=150mm]{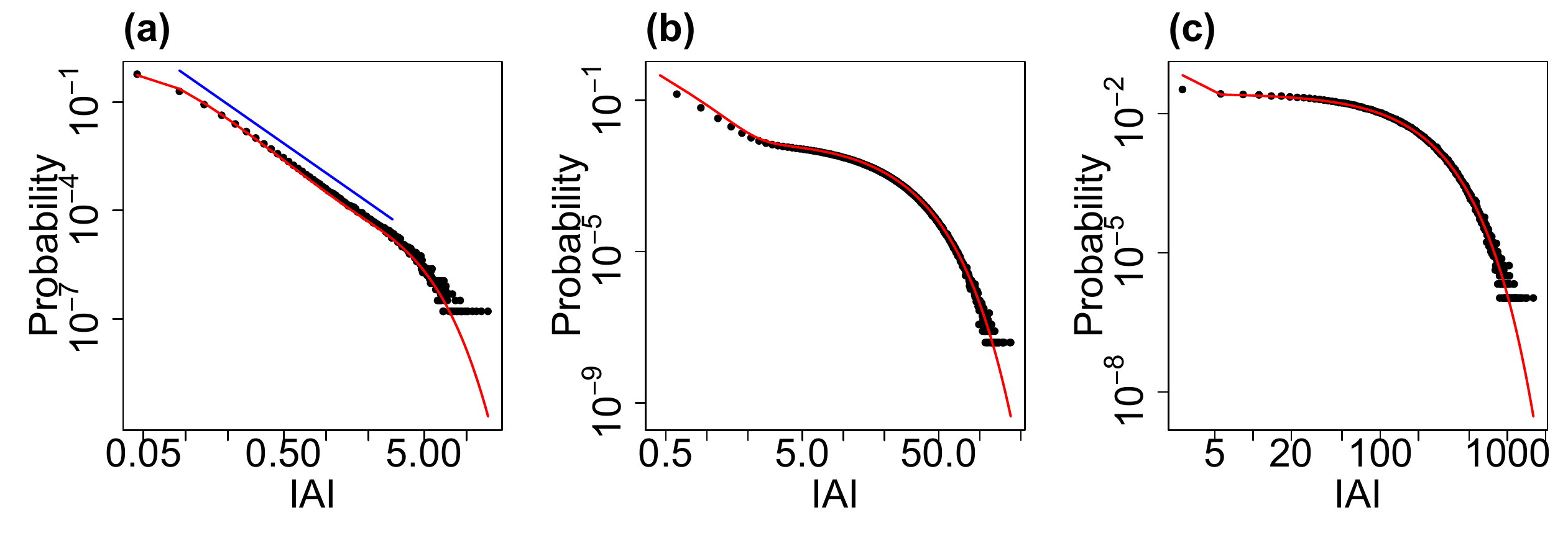}
\caption{\textbf{Distribution of IAIs for varying levels of the external input.} The theoretical (red) and simulated (black) IAI distributions with (a) $h=1/N$, (b) $h=0.1/N$ and (c) $h=0.01/N$. These distributions were for $N=800$ with $\alpha =w=1$, with a simulation length of $10^{4}$ seconds. The distributions from the simulated data are pooled from 10 simulations. The theoretical distributions were calculated up to the level of active neurons which occur with a cumulative probability of 0.9 (see main text). The blue line in plot (a) indicates a linear fit, i.e.\ a fitted power-law with an exponent of 2.71. }
\label{fig:IAIdists}
\end{figure}

\begin{figure}
\centering
	\includegraphics[width=50mm]{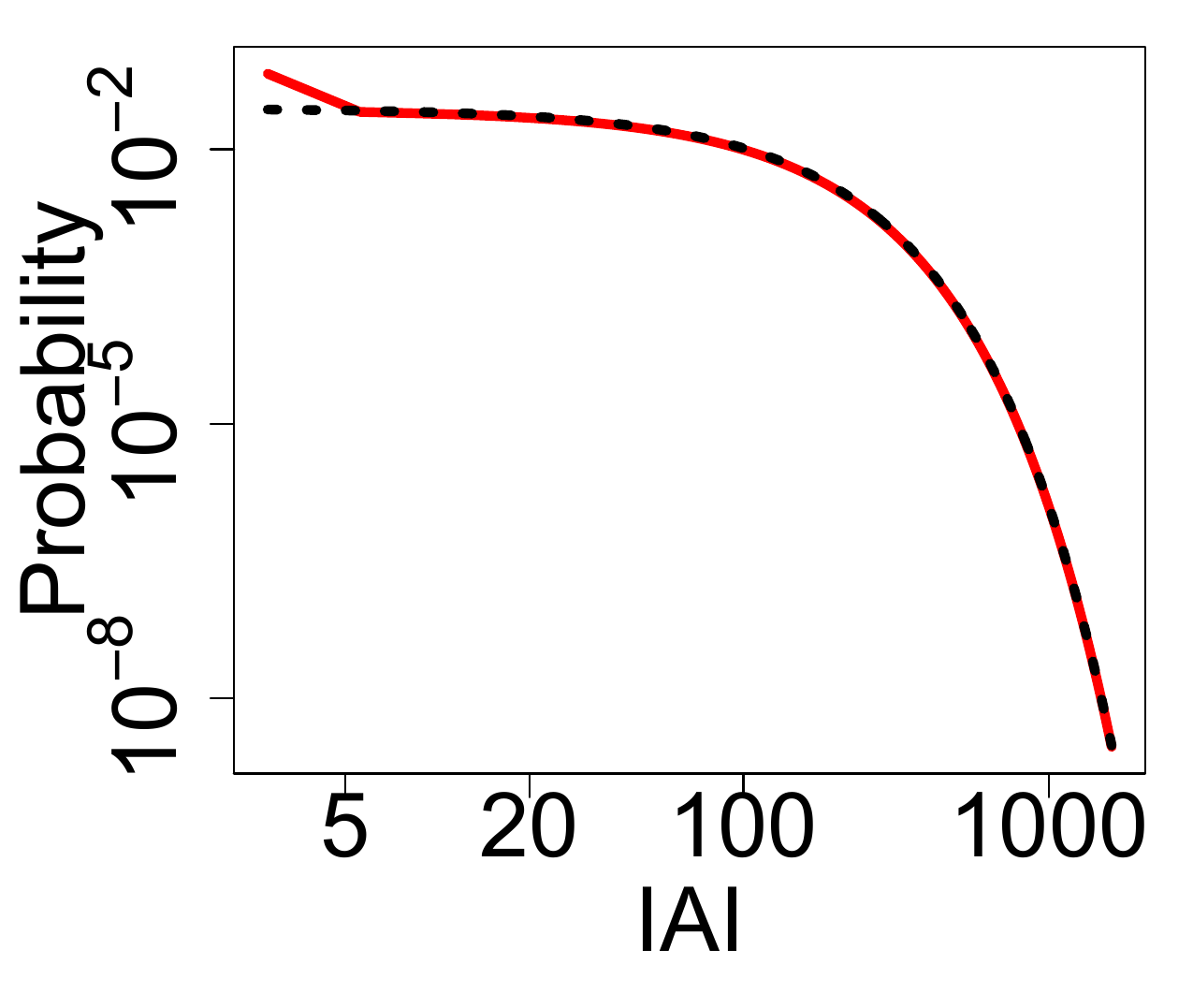}
\caption{\textbf{The IAI distribution with $h=0.01/N$ is well fitted by an exponential distribution.} The theoretical IAI distribution at the lowest level of the external input (shown in red, see also Fig.~\ref{fig:IAIdists}c) compared with the fitted exponential distribution (black dashed). The exponential is given by $y=0.028e^{-0.01x}$. This indicates that as the external input is decreased and the system approaches the critical regime the IAI distribution loses the scale-free behaviour seen at higher levels of the external input and is dominated by an exponential.}
\label{fig:IAIexponential}
\end{figure}

As an aside, note that due to the product in the hypoexponential (see equation~\ref{eq:hypoexp}) determination of the probabilities for large N can become computationally intractable. For simulations larger than with $N=50$ we therefore only determined the theoretical distribution up to a set level of the number of active neurons. We set the threshold level of the number of active neurons according to the probability distribution of starting from a particular number of active neurons (calculating the cumulative probability from zero active neurons), and sufficiently low so that the calculations were computationally viable. However, the theoretical distributions calculated using this threshold are still a good fit to the simulated data - Fig.~\ref{fig:IAIdists}.

\subsubsection*{Distributions of avalanche size and duration}

As we have shown, the theoretical distribution of IAIs can be calculated by assessing periods of consecutive active to quiescent transitions and single quiescent to active transitions. It is also possible to derive the distribution of consecutive quiescent to active transitions. However, if a period of active to quiescent transitions (a period without firing) has a duration less than the average time difference between two spikes then this interval does not separate an avalanche into two. Therefore, the distributions of number and length of consecutive quiescent to active transitions does not describe the distributions of avalanche size and duration - these distributions can also contain periods of active to quiescent transitions within two or more periods of quiescent to active transitions. Note that a period of active to quiescent transitions having a duration less than the average difference between consecutive spikes is not dependent on the number of active to quiescent transitions within the interval, as the length of each transition is drawn at random from an exponential distribution. It was therefore not possible for us to determine a theoretical distribution of avalanche size and duration using this approach.

\subsection*{Statistical comparison with a power-law distribution}

The influential paper by Clauset et al.~\cite{clauset} developed a model selection based methodology to determine whether empirical data are likely to be power-law distributed. This method has been used to assess physiological neuronal avalanches and the results have shown that the power-law hypothesis is not rejected for this data~\cite{klaus}. It is therefore of interest to determine whether this is also the case for the data from the model studied here.  Briefly, this method finds the best fit to a power-law of the distribution under study. The empirical data are then compared to distributions of the same size that are generated by randomly drawing values to follow the best-fit power-law distribution. A p-value is calculated as the proportion of times that the empirical data are a better fit to the power-law than the generated data (using the Kolmogorov-Smirnov test). As per Clauset et al.~\cite{clauset} the hypothesis (that the data come from a power-law) is rejected if the p-value is less than 0.1. As we have observed (Figs.~\ref{fig:distsize},~\ref{fig:IAIdists}), the distribution of avalanche sizes appears to exhibit partial scale-free behaviour for low levels of external input ($h=0.1/N,\ 0.01/N$) and the IAI distribution appears scale-free over a range of scales for $h=1/N$. As in the companion paper~\cite{taylor}, we fit a truncated power-law distribution up to an avalanche size of $x_{max}=\frac{9}{10}N$ in the case of avalanche size distributions. We fit a power-law distribution without truncation to the IAI distribution. Testing the entire avalanche size distributions (consisting of over 900,000 avalanches) yielded $p=0$ indicating that the hypothesis that the distribution follows a power-law should be rejected.  Similarly, taking the IAI distribution for $h=1/N$, testing the whole distribution of over 6,000,000 IAIs (note that there are more avalanches and therefore IAIs with larger $h$ due to the higher firing rate) yielded $p=0$. Testing instead the first 100,000 avalanches (a similar order of magnitude to the number of neuronal avalanches tested experimentally) with $h=0.1/N$ yielded $p=0.46$ indicating instead that the power-law hypothesis should not be rejected. Similarly, for $h=0.01/N$ testing the first 10,000 avalanches yielded $p=0.13$. These results are similar to those of the companion paper, where the power-law hypothesis was not rejected when the number of avalanches included in the distribution was of the same order as those tested experimentally, and are indicative of the partial scale-free behaviour of the system in proximity of the critical regime.
 
In the case of the IAI distribution testing the first 100,000 IAIs yielded $p=0.44$ indicating that a power-law is a good fit to the data. Given that in this case we know that the IAI distribution is not a power-law (but is in fact a weighted sum of hypoexponentials), it is interesting to note that the hypothesis that the data follow a power-law is not rejected when the number of data points is of the same order as that which have been tested experimentally, an observation that will be explained in the discussion. When the power-law hypothesis is not rejected, Clauset et al.~\cite{clauset} employ a model selection process to determine the best model for the data. We did not carry out this testing here (as, at least in the case of the IAI distribution, we already know what the distribution is) and it may be that such a process would suggest that a power-law is not the best fit to the data. However, the results here (and those of the companion paper) are indicative of the partial scale-free behaviour exhibited by the system in the region of the critical regime. 

\subsection*{Long-range temporal correlations}

As mentioned in the introduction, long-range temporal correlations are another possible signature of a system at (or near) a critical state and have also been observed in neurophysiological data~\cite{lk01,lk04,nikulin04,nikulin05,berthouze,smit}. It is therefore of interest to determine whether this finite-size neuronal system with external input displays LRTCs - given that it is in the region of a critical regime - and whether LRTCs relate to other signatures of criticality, i.e.\ the presence of partial scale-free behaviour in the data distributions themselves. The latter is of particular interest given that we have seen a change in distributions as the system approaches the critical regime. As with all simulations the level of external input is constant, i.e., it does not itself display LRTCs, and therefore we stress at the outset that any LRTCs present in the dynamics of the system would have to be intrinsic to the system. Furthermore, it is useful to remember that a power-law distribution of any data set does not imply that the data will exhibit LRTCs and vice-versa -- consider points drawn at random from a power-law distribution - such a data set would not exhibit LRTCs.

In neurophysiological data, LRTCs have been observed in fluctuations of oscillation amplitude (i.e.\ within continuous data)~\cite{lk01,lk04,nikulin04,nikulin05,berthouze,smit} and also in discrete burst activity in our recent analysis of the inter-event intervals of bursts of nested oscillations in EEG recordings of extremely preterm human neonates~\cite{hartley}. Moreover, LRTCs in discrete data have previously been investigated by Peng et al.~\cite{peng95} and a number of other authors, for example~\cite{toweill,castiglioni,ho}, in their analysis of inter-heartbeat intervals. As the data from the model analysed here result in discrete avalanche activity, we adopt the approach used in these previous studies of LRTCs in discrete data and examine LRTCs in waiting times, i.e.\, in IAIs.

We assessed the presence of LRTCs in IAIs through estimating the Hurst exponent, $H$, which describes the degree of self-similarity within the data. A Hurst exponent of $H\approx0.5$ indicates that there are no correlations in the data or short-range correlations only, for example a white noise process, whereas a Hurst exponent of $0.5<H<1.0$ indicates LRTCs in the data. Additionally, an exponent of 1 corresponds to $1/f$ noise~\cite{peng95}. We estimated the Hurst exponent using detrended fluctuation analysis (DFA) -- an approach that has been shown to produce more accurate estimates of the Hurst exponent than some other approaches~\cite{taqqu} and has been used previously to assess the presence of LRTCs in neurophysiological data sets~\cite{lk01,lk04,berthouze,hartley}. DFA is a graphical method whereby the average root mean square fluctuations at a given box size are compared across different box sizes and the gradient of the line of best-fit is the estimate of the Hurst exponent (for more detail see Peng et al.~\cite{peng95,peng94}). We used a minimum box size of 5, with 50 box sizes linearly spaced on a logarithmic scale up to a maximum box size of $1/10$ of the length of the IAI sequence~\cite{hu}. Calculations were carried out using the MATLAB code of McSharry~\cite{mcsharry}. 

Fig.~\ref{fig:dfaIAI} shows example DFA plots for IAIs from three simulations with $\alpha=w=1,\ h=1/N$. It is important to notice from these plots that there is not a single linear trend across all box sizes. Hu et al.~\cite{hu} discussed the importance of identifying crossover points - box sizes at which there is a change in the linear fit of the data - within DFA plots. Failure to examine these trends leads to erroneous estimates of the Hurst exponents. A single linear fit across all the points would give an estimate of the Hurst exponent for that sequence. However, crossover points indicate that the same correlations (i.e.\ temporal behaviour) do not extend across the whole sequence. In the DFA plots here there are in fact three regions, each with a different linear trend, between two crossover points. The best-fit to the data by three linear regions was found using the nonlinear regression function `nlinfit' in the MATLAB environment, therefore determining the crossover points. In Fig.~\ref{fig:dfaIAI}(a), the Hurst exponent (slope of the line) of the first two regions (at smaller box sizes) are 0.83 and 0.62 respectively - exponents which indicate the presence of LRTCs within the data. However, the third region across the largest box sizes has an exponent of 0.51 indicating that there are no correlations in the data. This change in the exponents therefore suggests that the correlations observed in the data at small box sizes do not extend across the entire sequence length.

\begin{figure}
\centering
	\includegraphics[width=150mm]{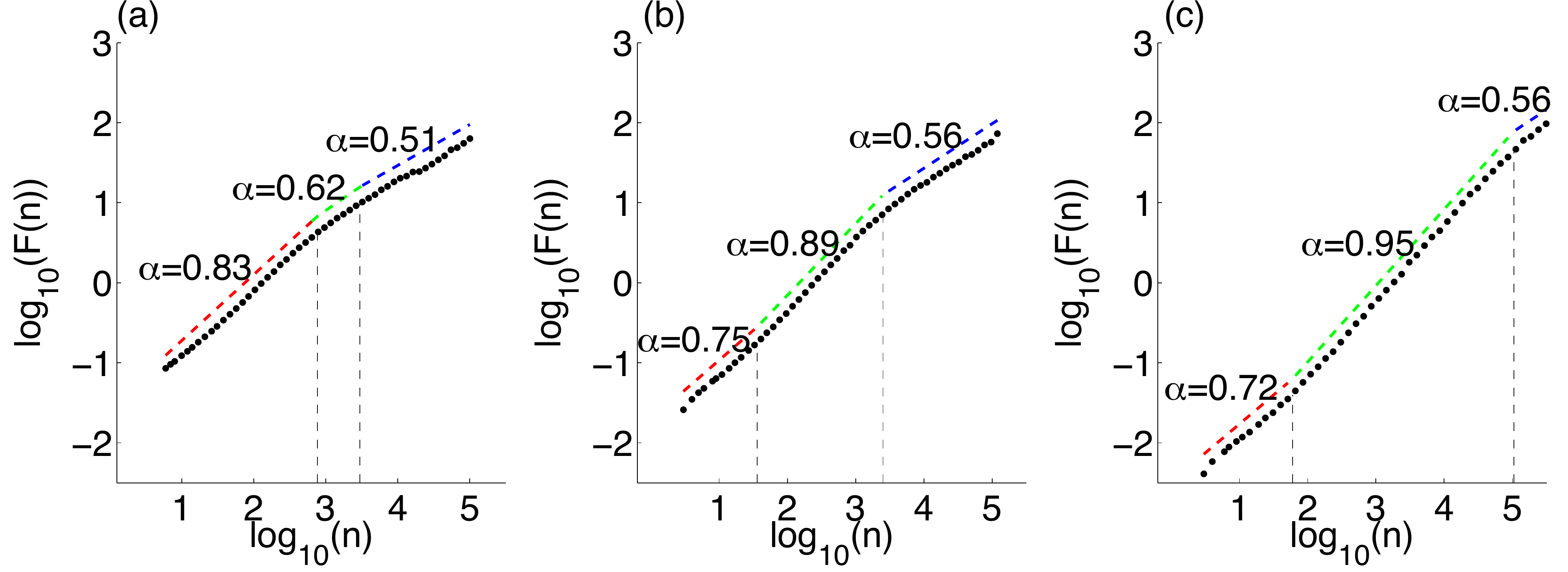}
\caption{\textbf{DFA plot examining the presence of temporal correlations in IAIs.} Plot of the average fluctuations $F(n)$ against box size $n$ for  IAIs from simulations with $\alpha =w=1,\ h=1/N$ and (a) $N=800$, (b) $N=3200$ and (c) $N=172800$. The data is best fit not by a single linear trend but by three lines (red, green, blue) between two crossover points (dashed black lines). For smaller box sizes the Hurst exponents (slope of the line - as annotated next to the individual lines: the DFA exponent $\alpha$) indicates that correlations extend across these regions. However, for larger box sizes the exponents are closer to 0.5 suggesting that the correlations do not extend across these larger box sizes. With a larger system size (c) the upper crossover point increases to larger box sizes. With increasing system size the system approaches the critical regime - (a) $\lambda=-0.07$, (b) $\lambda=-0.04$ and (c) $\lambda=-0.005$.}
\label{fig:dfaIAI}
\end{figure}

%max and min box size etc - also in conn form paper

When examining the presence of LRTCs it is standard practice to compare the exponent of the actual data to the exponent of the data randomly shuffled~\cite{lk01}. Shuffling the data should destroy any correlations present and therefore the exponent of the shuffled data is expected to be approximately 0.5. We compared the original sequence (whose DFA plot is shown in Fig.~\ref{fig:dfaIAI}) with 500 shuffled sequences. As expected, the DFA plots for the shuffled sequences (data not shown) did not exhibit crossover points and showed a mean exponent for the shuffled sequences of 0.50 with a range of 0.48-0.52. The fact that the exponents of the original sequence (at smaller box sizes) do not fall within the distribution of exponents for the shuffled sequences therefore demonstrates that the original sequence exhibits complex temporal ordering with correlations that extend across a range of box sizes (up to the upper crossover).

\subsubsection*{Increasing the system size}

As noted previously (see Fig.~\ref{fig:eigenvalueNh}), as $N\rightarrow \infty$ the eigenvalue of the system $\lambda \rightarrow 0$, i.e., the system approaches the critical regime with increasing system size. We might expect that as the system approaches the critical regime it is more likely to exhibit signatures of criticality and therefore that LRTCs would extend to larger box sizes as the system size is increased. We therefore investigated whether there was a change in the temporal correlations of the IAIs with system size, while maintaining all other parameters including $h=1/N$. For all system sizes investigated the DFA plots displayed three regions with different linear trends, as was discussed above. Fig.~\ref{fig:dfaIAI} shows example DFA plots for the IAIs of three simulations with the smallest and largest system sizes examined. It was observed that the pattern of the exponents in each of the cases remained the same - with the two lower regions having exponents indicative of LRTCs, while the exponent across the largest box sizes is closer to 0.5. Additionally, we found that the location of the upper crossover increased with system size. Fig.~\ref{fig:changeN} shows the location of the upper crossover (the crossover at the higher box regions) for different system sizes (from N=3200 to N=172800) normalised with respect to the largest box size. We did not find a change, other than small fluctuations, in the exponents themselves for any of the three regions for all system sizes. Namely, across all system sizes considered, the mean exponent across the smallest box sizes (up to the first crossover point) was 0.73 with a range of 0.70 - 0.76. It was 0.95 with a range of 0.89 - 0.99 (close to an exponent of 1 which would indicate 1/f noise) between the first and second crossover points. The largest variation in exponents was for the region above the upper crossover point with an average exponent of 0.59 and a range of 0.46 - 0.73; on average this indicates that temporal correlations do not extend beyond the upper crossover. Thus, overall, as the system size is increased the temporal correlations extend across larger box sizes. This is consistent with the idea that, when the system reaches the critical regime, LRTCs could extend to infinite length (i.e., all possible box sizes) in the limit of system size. 

\begin{figure}
\centering
	\includegraphics[width=160mm]{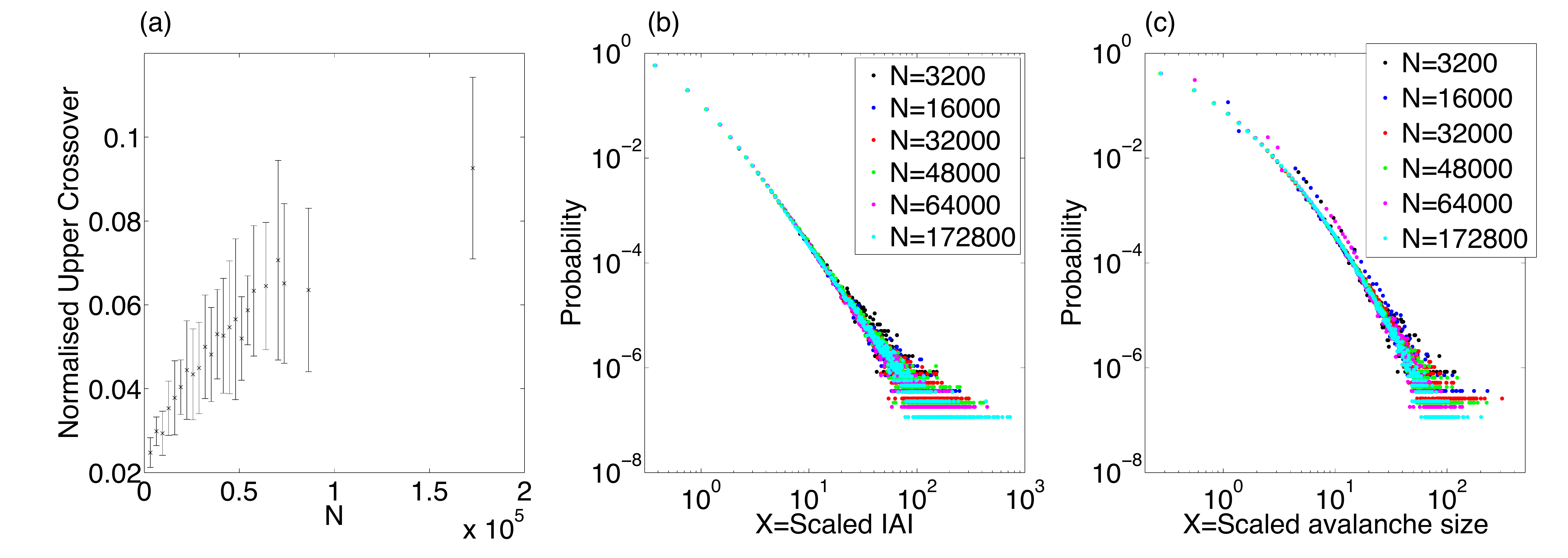}
\caption{\textbf{Changes with increasing system size.} (a) The normalised (with respect to the largest box size) upper crossover box size increases with system size. The plot shows the average value across 10 simulations at the same system size and error bars indicate the standard deviation. (b) The IAI distributions and (c) the distributions of avalanche size for 6 different system sizes, scaled with respect to the mean IAI or avalanche size of each distribution respectively. For all simulations $\alpha =w=1,\ h=1/N$.}
\label{fig:changeN}
\end{figure}

Next we considered whether the distributions of IAIs and avalanche size themselves changed with system size and whether the change in the correlation length observed above was reflected in a change in the distributions. Fig.~\ref{fig:changeN} also shows the IAI and avalanche size distributions for different network sizes. In both cases, the distributions for different system sizes show only small changes which can be accounted for by noise. Thus, as the system approaches the critical regime through increasing the system size there does not appear to be a change in the distributions despite the change in the temporal correlations. Moreover, LRTCs are present in the data but the distribution of avalanche sizes does not exhibit scale-free behaviour, i.e., these markers of criticality do not occur simultaneously in this case. By contrast, through approaching the critical regime by lowering the external input we have shown that the avalanches are more distinct and the distribution of avalanche sizes exhibits partial scale-free behaviour. 

\subsubsection*{The effect on LRTCs of decreasing the level of external input}

We also examined the DFA exponents at lower levels of external input ($h=0.1/N$ and $h=0.01/N$). In both cases there were no crossover points with a single linear trend across all box sizes (data not shown). The exponents were 0.50 (range 0.49-0.51, across 10 simulations with $N=800$) and 0.56 (range 0.55-0.57) for $h=0.01/N$ and $h=0.1/N$ respectively. Thus, at the lowest level of external input the IAIs do not exhibit LRTCs and there is a slight increase in the exponent as the external input increases. This suggests that as the system approaches the critical regime through a decrease in the external input the temporal correlations are lost. Thus, the existence of LRTCs as the system approaches the critical regime is dependent on how the critical regime is approached, namely, approaching the critical regime through increasing the system size extends the temporal correlations whereas decreasing the external input leads to a loss of long-range correlations. In addition, this signature of criticality is independent from the other marker we investigated -- the presence of scale-free behaviour in the avalanche size distribution. Considering avalanche size and duration, scale-free behaviour is present at the lowest level of external input, when LRTCs are lost. Thus, we find that markers of criticality are not only dependent on the region around the critical regime but also may not be present for the same parameter set.

\section*{Discussion}

This paper specifically examined a finite-size neuronal system without a separation of timescales between external input and the avalanches themselves. By analytically tuning the system to be in the region of a critical regime we were able to examine the type of dynamics displayed by such a system and to investigate whether the dynamics display signatures of criticality. 
In summary, we have shown that:
\begin{enumerate}
\item As the system approaches the critical regime through a reduction in the external input the avalanches become more distinct and the distribution of avalanche sizes displays scale-free behaviour.
\item With $h=1/N$ the IAIs exhibit temporal correlations which extend across a range of bin sizes to an upper crossover. As the system approaches the critical regime through increasing the system size the length of the temporal correlations is extended across a wider range of bin widths. These correlations (one noted signature of a critical system) are observed despite the fact that the distribution of avalanche sizes does not exhibit scale-free behaviour and does not change with the increase in system size. These temporal correlations are lost if the critical regime is instead approached through reducing the external input.
\item The distribution of IAIs was theoretically derived and was shown to be a weighted sum of hypoexponentials. However, for $h=1/N$ (when the number of avalanches considered was of the same order as those tested experimentally) the hypothesis that the IAI distribution follows a power-law was not rejected by statistical testing indicating the scale-free nature of the distribution at this level of the external input.
\end{enumerate}

\subsection*{Validity of the model}

The model considered in this paper was a highly simplified neuronal system with a number of assumptions, such as equally weighted synapses and continuous constant external input. These assumptions were necessary in order to analytically tune the system to be in the region of a critical regime. Therefore, while this should not be taken as an accurate model of a real neuronal system it is important that we first consider models such as this, examining markers of criticality, which will then aid our understanding when building on this work with more complex models. This paper opens the way for future work examining the role of external input on signatures of criticality and the importance of the region of parameter space on network dynamics. Future work should also investigate the effect of topology on the dynamics~\cite{larremore,sporns} and the effect of external input with different temporal and spatial characteristics.

\subsubsection*{A purely excitatory network}

The synaptic connections investigated in this model were purely excitatory. This not only simplifies the model for analytical investigations but is also of interest from a neurological perspective in terms of early brain development. Before birth, GABA is thought to have a depolarising effect on postsynaptic neurons and it is not until the nervous system reaches a more mature state that this neurotransmitter becomes inhibitory~\cite{cherubini,benari}. While \textit{presynaptic} inhibition is thought to be present at all developmental stages~\cite{holmes} this effect can be considered to be taken into account in the model by the fact that neurons cannot re-fire until they have returned to the quiescent state. We have recently shown that EEG recordings from very preterm infants (when GABA is still thought to be purely excitatory) exhibit LRTCs in the temporal occurrence of bursts of activity~\cite{hartley}. The model studied here may be a candidate mechanism for the generation of this temporal patterning in the discontinuous activity of the developing brain. Moreover, it is interesting to note that despite the fact that the system has purely excitatory postsynaptic connections and input, for these parameter regions, the model does not exhibit runaway excitation (saturation) but is able to maintain stable dynamics through the `balance' of individual neuronal dynamics resulting from a trade-off between the rates at which neurons become active and quiescent. Indeed, while a number of authors have suggested that a balance of excitation and inhibition in neuronal networks leads to critical behaviour~\cite{shew}, the work here and in the companion paper shows that excitatory networks can display the same behaviour. It can be speculated that this type of balanced activity in the region of a critical regime might be a way in which the brain avoids (for the most part) epileptic behaviour during early development.  

\subsubsection*{The activation function}

Here we used a linear activation function for the transition of neurons from the quiescent to the active states. However, physiologically neurons behave more like a saturating function. The linear activation function used here was chosen so as to be analytically tractable and is also equivalent to a saturating function at low input levels. However, considering instead a saturating function (see Appendix 3) we found the dynamics in the region of the critical regime to show similar behaviour to the system with the linear activation function. 

With both the linear and saturating activation functions, the critical regime can only be reached exactly in the absence of external input. A positive external input therefore drives the system away from the critical regime. However, with a quadratic activation function (see Appendix 3) the system does have a critical fixed point even with a positive external input and it can be tuned directly to this regime. With such an activation function the dynamics do not appear to exhibit burst like behaviour, however, analysis shows that the activity fluctuates about the critical regime in an `avalanche-like' manner. Thus, while a quadratic function does not best describe activation in a neuronal network, we can emphasise our conclusion that signatures of criticality are not universal and can be examined only in relation to the specific critical regime of the system (see Appendix 3). 

\subsubsection*{The binning approach}

As described previously, the binning method separated avalanches where the time difference between consecutive spikes was greater than the average time difference between consecutive spikes across the entire simulation. This was the approach taken by Benayoun et al.~\cite{benayoun}. However, it is worth noting that this is a slightly different approach to the method first proposed by Beggs and Plenz~\cite{bp1,bp2} to separate neuronal avalanches. In their analysis neuronal firing is distributed into bins of width of the average time difference between consecutive spikes ($\delta t$) and firing is separated into avalanches by bins in which no firing occurs. Thus, two spikes may be separated by more than the average time difference $\delta t$ but still be considered part of the same avalanche if they fell within consecutive bins. Our theoretical derivation of the IAI distribution relied on the fact that all consecutive active to quiescent transitions or single quiescent to active transitions with a length greater than the average time between two spikes is an IAI. This would not be the case if the alternative (Beggs and Plenz) binning approach was used to determine avalanches. If this alternative approach was used the distributions of consecutive active to quiescent transitions and single quiescent to active transitions would be the same, but transitions of length slightly greater than or equal to the average time between consecutive spikes (in fact up to twice this average) may or may not form part of the IAI distribution depending on the exact binning. It is also important to note that with the binning method used here, even with dense neuronal firing (which occurs if the external input is increased from the levels studied here), it is always possible to separate the dynamics into `avalanches' since there is always an average time between consecutive spikes. 

Additionally, both these binning approaches differ from that used in non-driven systems such as the classical sand-pile model~\cite{bak} and the system investigated in the companion paper to this article~\cite{taylor}. In those models an avalanche consists of all firings until the system returns to the fully quiescent state. This means that although the system may have a long period without firing -- during which neurons switch to the inactive state --  activity before and after this period will not be considered as separate avalanches even if the period in-between exceeds the average difference between consecutive spikes. Future work is needed to fully investigate how differences in the definitions of avalanche affect the distributions of size, duration and IAIs and care needs to be taken when interpreting the results from these different approaches.

\subsubsection*{Validity of DFA and the investigation of LRTCs}

DFA is one method by which to estimate the Hurst exponent and was chosen here as it has been shown to be an accurate estimate~\cite{taqqu}. Moreover, it is a graphical approach and so can be used to check for crossover points~\cite{hu} -- see also the recently proposed ML-DFA~\cite{mlDFA} whereby the existence of crossover behaviour can be rigorously tested. As the Hurst exponent can only be estimated it is is considered best practice to check the consistency of the exponents using two methods~\cite{gao}. However, as non-graphical methods only give single numerical values they cannot be interpreted when crossover behaviour exists. Given that there were crossover points we only considered DFA with this analysis. 

Crossover points within a DFA plot have been shown to exist when the same correlations do not extend across the whole data sequence in analytically constructed data~\cite{hu}. It is important to correctly interpret these crossover points. Being box sizes, they define sequence lengths, e.g., a box size of 10 indicates detrending across a sequence of 10 consecutive IAIs. Note that as the IAIs are of variable length the box size does not specify a particular simulation time but merely a number of events. Future investigation is therefore needed to determine the relationship between the model and its crossover points.

Correlations extended only across a range of box sizes, with this range extending as the system size increased and the system approached the critical regime. It appears that correlations could potentially extend across an arbitrarily large box size in the limit of system size. Thus, as the critical regime is approached in this way, this signature of a critical system emerges. LRTCs have been demonstrated previously in discrete neurophysiological data, in the waiting times of burst activity in cultures~\cite{segev} and in the bursts of activity recorded using EEG in very preterm human neonates~\cite{hartley}. To our knowledge, waiting times of neuronal avalanches have yet to be examined in this manner. However, such a study would provide an additional link between studies on the neuronal scale and studies on a wider network scale for which LRTCs have been observed in the fluctuations of oscillation amplitude. Palva and colleagues demonstrated strong correlations between power-law exponents of avalanche size distributions and exponents of LRTCs in fluctuations of oscillation amplitude in human MEG recordings~\cite{palva}. Recent computational work also demonstrated a link between neuronal avalanches on the one scale and LRTCs on a wider temporal scale and the authors called for future work in this area~\cite{poil}. However, these authors did not investigate LRTCs in the waiting times of the avalanches themselves. Interestingly, in our model, LRTCs were observed when $h=1/N$ but not for lower levels of external input. Thus, they were not observed when the avalanche size distribution exhibited scale-free behaviour -- the type of distribution observed for avalanches recorded \textit{in vivo} and \textit{in vitro}~\cite{bp1,bp2,petermann}. It would therefore also be interesting to assess whether altering the driving force experimentally \textit{in vitro} would lead to the types of dynamics (LRTCs) observed here.

\subsection*{Partial scale-free behaviour in avalanche size}

Statistical testing of the avalanche size distribution (with $h=0.1/N,\ 0.01/N$) did not reject the hypothesis that the distribution followed a power-law when the number of points within the distribution was of the order of the number of avalanches recorded in the experimental setting. Only with larger numbers of avalanches was the hypothesis that the distribution is a power-law rejected. This is to be expected, as has been discussed by Klaus and Plenz~\cite{klaus}. When a distribution deviates from the expected distribution by more than noise from sampling then given a large enough number of samples the power-law hypothesis will eventually be rejected. The fact that the power-law hypothesis was not rejected for lower numbers of avalanches demonstrates the partial scale-free behaviour of the system in the region of the critical regime. Further, it highlights the fact that stringent statistical testing, such as this, with high sampling may lead to rejecting the power-law hypothesis and so rejecting the criticality hypothesis even when the system is critical.

\subsection*{Waiting times}

In addition to increasing the physiological realism of the model, investigating the system with continuous external input also has the advantage of producing waiting times (termed IAIs throughout). In the companion paper the simple reseeding of the network with a neuron set to the active state implied that there was no waiting times between avalanches. Other authors have reseeded by increasing the membrane potential but stipulated that neurons must reach a threshold for them to become active (and a new avalanche to start)~\cite{bak,levina}. This does lead to waiting times, however, these are not the same as the waiting times investigated in this model which are intrinsic to the network dynamics rather than as a result of an explicit separation of timescales. 

Recent work by Lombardi and colleagues~\cite{lombardi} showed that the waiting times between neuronal avalanches recorded \textit{in vitro} have a distribution with an initial power-law regime. The authors suggest that the shape of the distribution relates to up and down states within the network (which exhibit critical and subcritical dynamics respectively) and are able to reproduce the non-monotonic waiting time distribution in a computational model in which neurons switch between up and down states depending on short-term firing history. Interestingly, the distribution they observe is similar to the IAI distribution for the system with $h=0.1/N$, see Fig.~\ref{fig:IAIdists}(b), which also has a scale-free initial regime albeit over a shorter range to that presented by Lombardi et al. It is therefore possible that the waiting time distribution observed experimentally fits with the model constructed here. It would be interesting to investigate whether a change in input to the network \textit{in vitro} alters the distribution in a similar way to those distributions seen in Fig.~\ref{fig:IAIdists}.

Additionally, for different parameter ranges different distributions were observed, in the IAI distribution as well as the distributions of avalanche size and duration. This leads us to the important conclusion that power-law distributions will not necessarily be displayed by systems in the region of a critical regime. Therefore, this work suggests that the absence of a power-law in experimental data should not necessarily be taken to conclude that the system does not lie in the region of a critical regime. This was also seen in the companion paper where it was shown that despite being analytically tuned to the critical state (in absence of external input) the avalanche size distribution was not a power-law although it did exhibit partial scale-free behaviour. The fact that the system may not exhibit power-laws when close to (or at) the critical regime is an important finding given that the system is finite-size as will be the case in an experimental setting. This highlights the necessity of examining other markers of criticality before conclusions about the critical nature of a system can be drawn.

\subsection*{Dynamic range and power-laws}

Coinciding with results from previous authors~\cite{kinouchi,larremore} we showed that the system exhibits optimal dynamic range when the branching parameter is equal to one. When calculating the dynamic range of a system, we emphasised that this value was dependent on the critical state of the system calculated when there was no external input.  We have shown that tuning a system to this critical point but then driving it with different levels of external input has considerable effect on the distribution of avalanche sizes.  For non-zero $h$ the corresponding ODE would, in the strictest sense, not be considered critical.  Importantly, however, tuning to the critical point of the system with zero external input, maximises the dynamic range.

Dehghani and colleagues~\cite{dehghani} showed that \textit{in vivo} (against the results of Petermann et al.~\cite{petermann} and Hahn et al.~\cite{hahn}) avalanches were not well approximated by power-laws, but were more likely to approach exponential distributions.  They contrast this with the evidence that the brain is operating at criticality from \textit{in vitro} studies \cite{bp1,friedman} where avalanches are well approximated by power-laws.  Here we argue that external input and functional benefits~\cite{shewreview} such as dynamic range, information transmission and information capacity, provide an interesting possibility as to the reason why \textit{in vivo} and \textit{in vitro} studies could potentially give different results.  The critical brain hypothesis demands that in isolation from its natural surroundings (\textit{in vitro}) and whilst having no external influences acting upon it (akin to the model with $h=0$ we studied in the companion paper~\cite{taylor}), a culture should exhibit signs that it is tuned to criticality (i.e.\ avalanches that are well approximated by power-laws).  However, when observed \textit{in vivo}, and thus with external inputs acting upon it, a critical brain may no longer exhibit avalanches approximated by power-laws but still optimise functional benefits such as the dynamic range and information transmission~\cite{shewreview}.   In our model we have shown that tuning the parameters to the critical regime does indeed maximise the dynamic range, but it is the level of external input that dictates whether the avalanche distributions exhibit partial scale-free behaviour.  For this reason, avalanches recorded \textit{in vivo} that lacked  a power-law distribution would not be in contradiction with the criticality hypothesis but rather an expected result. This further supports our suggestion in the companion paper~\cite{taylor} that future work should shift its focus away from characterising avalanche distributions and towards more appropriate metrics.

\subsection*{Two routes to criticality}

In this paper we examined two different parameter changes such that the system approaches the critical state: increasing the system size and lowering the overall level of the external input. Despite the fact that in both cases the critical regime is approached, the dynamics and the signatures of criticality observed are different. With increasing system size the temporal correlations extend across a wider range. However, the distributions of the avalanche characteristics remain the same and the distribution of avalanche size does not exhibit scale-free behaviour. By contrast, for lower overall levels of the external input the distributions of avalanche size and duration do exhibit partial scale-free behaviour. However, in this case as the critical regime is approached the temporal correlations in the avalanches are lost. At these lower levels of the external input we also observe a greater separation of the avalanches suggesting that the avalanches have less of an influence on each other which would explain this loss of LRTCs. Thus, as the system approaches the critical state in two different regions of the parameter space the dynamical properties of the system are very different. Significantly, this implies that not just the critical state alone but the region around the critical regime is an important factor in the system's dynamics.

In conclusion, we have shown here and in the companion paper that in a finite-size neuronal system in the region of a critical regime the distributions of avalanche attributes need not be a power-law. The current assumption in the literature is that power-law dynamics imply criticality and vice versa that systems without power-law dynamics are not in the region of a critical regime, however, the results here suggest that this assumption need not be true. Moreover, we found that long-range temporal correlations and scale-free distributions are not dependent on proximity to the critical regime alone but on the region of the parameter space. The results further highlight the need for future work examining the type of dynamics we might expect from such systems.

\section*{Appendix 1: Dynamic range}

Whilst~\cite{kinouchi}~and~\cite{larremore} consider a discrete model where multiple events can happen per time step, here we show analytically that our continuous model will exhibit the same maximisation of the dynamic range when $R_0 = 1$. We use the calculation of $R_0$ for a system where there is no external input ($h=0$) and thus $R_0 = w/\alpha$.

We begin by defining (as in Kinouchi and Copelli~\cite{kinouchi})  $F_{max}(R_0)$ as the saturation level of neurons in a network assuming a large external input $h$.  For our model $F_{max}(R_0)=N$ for all $R_0$.  Similarly we define $F_{0}(R_0)$ as the steady state solution of the mean field ODE for the system when there is zero external input, i.e.
\begin{align*}
\frac{dA}{dt} 	&= \left(\frac{wA}{N}+h\right)\left(N-A\right)- \alpha A \\
			&= \frac{wA}{N}(N-A )- \alpha A.
\end{align*}
Therefore, solving this we have that
\begin{align*}
F_0(R_0) = \begin{cases} 0 &\mbox{if } R_0 \leq 1 \\
								N\left(1-\frac{\alpha}{w}\right) & \mbox{if } R_0 >1. \end{cases}
\end{align*}
Additionally let $F_{x}(R_0) = F_{0}(R_0) + x\left[F_{max}(R_0) - F_{0}(R_0)\right]$ giving
\begin{align*}
F_x(R_0) = \begin{cases} Nx &\mbox{if } R_0 \leq 1 \\
								N\left[1-\frac{\alpha}{w}\left(1-x\right)\right] & \mbox{if } R_0 >1. \end{cases}
\end{align*}
Finally, let $A(\sigma,y)$ be the number of active neurons at the steady state in a regime where $R_0 = \sigma$ and $h = y$ (where $\sigma$ and $y$ are dummy variables and $h$ is the external input), then the dynamic range $\Delta(R_0)$ is defined (similarly to~\cite{kinouchi}) as: 
\begin{align*}
\Delta(R_0) = \frac{h_{0.9}}{h_{0.1}},
\end{align*}
where
\begin{align*}
h_{0.1} \text{ is the level of external input such that } A(R_0,h_{0.1}) &= F_{0.1}(R_0) = F_{0.1} \\
\text{and }h_{0.9} \text{ is the level of the external input such that } A(R_0,h_{0.9}) &= F_{0.9}(R_0) = F_{0.9}.
\end{align*}
We note that in \cite{kinouchi,larremore}, the logarithm of this is taken but as the logarithm is an increasing function it is unnecessary to scale in this way for the result we obtain.  Whilst using $F_{0.1}$ and $F_{0.9}$ is the standard for calculating the dynamic range these values are somewhat arbitrary~\cite{kinouchi} and can be generalised to $k_1$ and $k_2$ respectively.  To calculate the dynamic range analytically we consider the two regimes of $R_0$, firstly $R_0 \leq1$ and secondly $R_0 > 1$.

\subsection*{$R_0 \leq 1$}
Here the steady state is given by 
\begin{align*}
&\left(\frac{wF_{k}}{N}+h_k\right)\left(N-F_{k}\right) - \alpha F_{k} = 0 \\
\Rightarrow &\left(wk+h_k\right)(N-Nk) - \alpha N k = 0 \\
\Rightarrow &h_k = \frac{\alpha k}{1-k} - wk,
\end{align*}
thus
\begin{align*}
\Delta = \frac{h_{k_2}}{h_{k_1}} &=\frac{\left[\alpha k_2 - wk_2(1-k_2)\right]\left(1-k_1\right)}{\left[\alpha k_1 - wk_1(1-k_1)\right]\left(1-k_2\right)}\\
&= \frac{k_2(1-k_1)\left[1-R_0(1-k_2)\right]}{k_1(1-k_2)\left[1-R_0(1-k_1)\right]}
\end{align*}

\subsection*{$R_0 > 1$}
Here the steady state is given by 
\begin{align*}
&\left(\frac{wF_{k}}{N}+h_k\right)\left(N-F_{k}\right) - \alpha F_{k} = 0 \\
\Rightarrow &\left[w\left(1-\frac{\alpha}{w}\left(1-k\right)\right)+h_k\right]\left[\frac{N\alpha}{w}\left(1-k\right)\right] - \alpha N\left[1-\frac{\alpha}{w}\left(1-k\right)\right]=0 \\
\Rightarrow &h_k = \frac{k}{1-k}\left(w-\alpha+\alpha k\right)
\end{align*}
thus
\begin{align*}
\Delta = \frac{h_{k_2}}{h_{k_1}} &= \frac{k_2(1-k_1)(w-\alpha + \alpha k_2)}{k_1(1-k_2)(w-\alpha + \alpha k_1)} \\
&= \frac{k_2(1-k_1)(R_0-1 + k_2)}{k_1(1-k_2)(R_0-1 +  k_1)} 
\end{align*}

\subsection*{Maximum of $\Delta(R_0)$}
Calculating the derivative of $\Delta(R_0)$ we find that if $0 <k_1 <k_2 < 1$, then for $R_0 \leq 1$, $\frac{d \Delta}{d R_0} > 0$, whilst for  $R_0 > 1$, $\frac{d \Delta}{d R_0} < 0$.  Thus the maximum of $\Delta(R_0)$ is achieved for $R_0 = 1$.  It is worth noting that $\Delta(R_0)$ is independent of N and only depends on the choice of $k_1$ and $k_2$.

\section*{Appendix 2: Driving the system from a subcritical and supercritical state}

Throughout the paper we have examined parameters such that the system is critical when there is no external input. In the presence of a small external input we therefore investigate driving the system in the region of this critical state. In the companion paper~\cite{taylor}, with no external input, we also investigated the system with subcritical and supercritical parameters. In this appendix we briefly examine the dynamics of the system as it is driven from these states by an external input.

Fig.~\ref{fig:subsupraster} shows raster plots of network firings when the system is driven from a subcritical and supercritical state with $h=1/N$. Compared with the critical case, see Fig.~\ref{fig:examplesims}(a), with the subcritical parameter set the bursts appear to be shorter and consist of fewer neurons firing. Conversely, in the supercritical case the bursts appear longer and consist of denser network firing. Fig.~\ref{fig:subsupIAIdists} shows the IAI distributions for the subcritical and supercritical parameters. As expected from the raster plots, the IAIs are longer in the subcritical case compared with the critical (Fig.~\ref{fig:IAIdists}(a)) and the supercritical. While the subcritical distribution appears to exhibit partial scale-free behaviour similar to the critical case, the supercritical distribution loses this appearance. The distributions from simulations are shown with the theoretical distribution calculated as previously described as a weighted sum of hypoexpontials.

\begin{figure}
\centering
	\includegraphics[width=150mm]{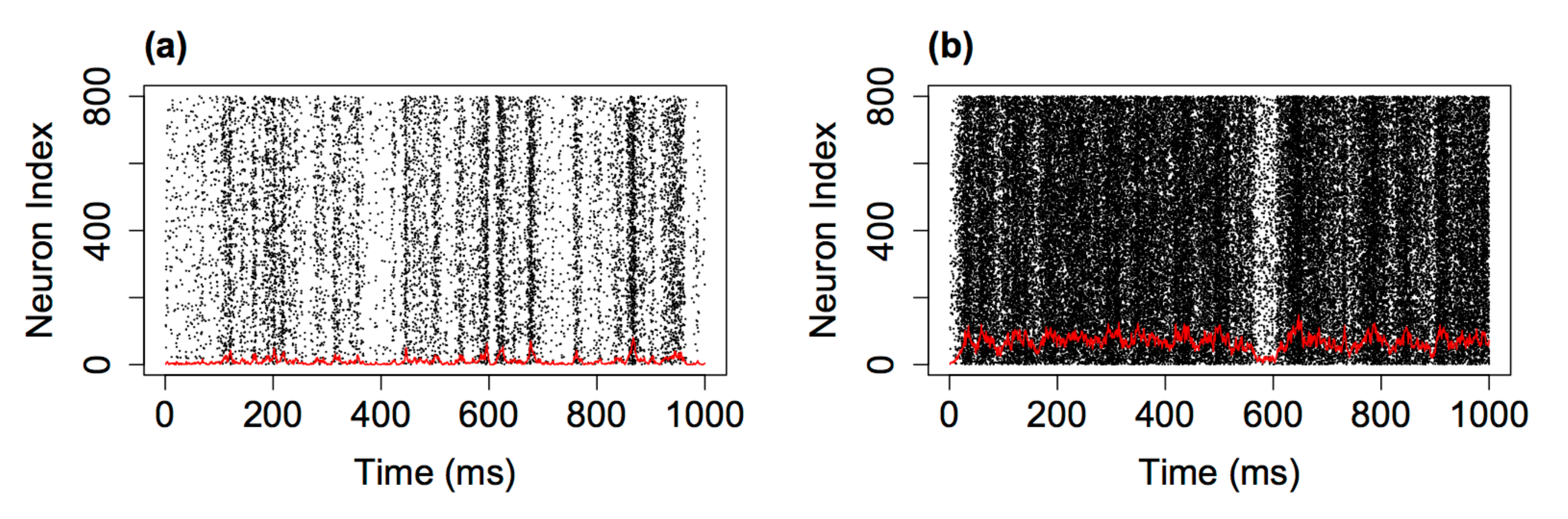}
\caption{\textbf{Raster plots of neuronal firing for the network driven from subcritical and supercritical states.} The network firing for (a) subcritical, $\alpha=1.1,\ w=1 \ \Rightarrow \ \lambda<0$ and (b) supercritical, $\alpha=0.9,\ w=1\ \Rightarrow \ \lambda>0$ parameter sets. Here we investigate the system with a small external input ($h=1/N$) which drives the system slightly away from these fixed points. The red line indicates the level of firing in 1 ms bins. The subcritical case appears to give rise to smaller bursts and the supercritical case leads to a greater level of firing and longer burst activity compared with the critical system (see Fig.~\ref{fig:examplesims}).}
\label{fig:subsupraster}
\end{figure}

\begin{figure}
\centering
	\includegraphics[width=150mm]{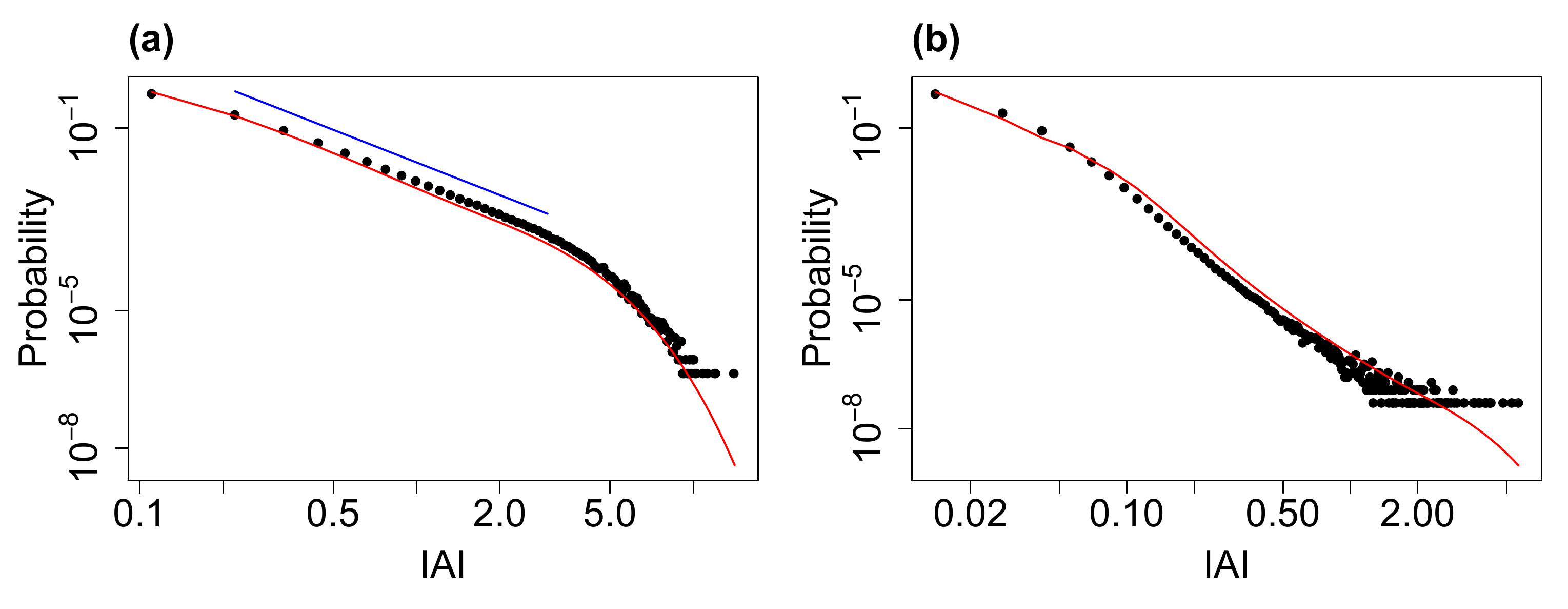}
\caption{\textbf{Distribution of IAIs for the system driven from subcritical and supercritical states.} The theoretical (red) and simulated (black) IAI distributions with (a) $\alpha=1.1$ (subcritical) and (b) $\alpha=0.9$ (supercritical) parameters. The blue line in (a) indicates a linear fit, i.e.\ a fitted power-law with an exponent of 2.37. These distributions are with $N=800,\ w=1,\ h=1/N$ and the theoretical distributions were calculated up to a level of initial active neurons which occur with a cumulative probability of 0.9 and 0.13 respectively (see main text).}
\label{fig:subsupIAIdists}
\end{figure}

Fig.~\ref{fig:subsupsizedists} shows the distributions of avalanche size and duration in the subcritical and supercritical cases. Contradicting what we would expect from the raster plots we find that the avalanche sizes are smaller (on average) in the supercritical system. In the companion paper we showed that the supercritical distribution (without the presence of external input) had an increased number of large avalanches compared with the distribution for the system at criticality. However, we do not find this here. As the firing with the supercritical parameters is relatively dense we believe that this highlights a limitation with the binning method in this case. We suggest that future research should focus on how binning can influence avalanche distributions.

\begin{figure}
\centering
	\includegraphics[width=150mm]{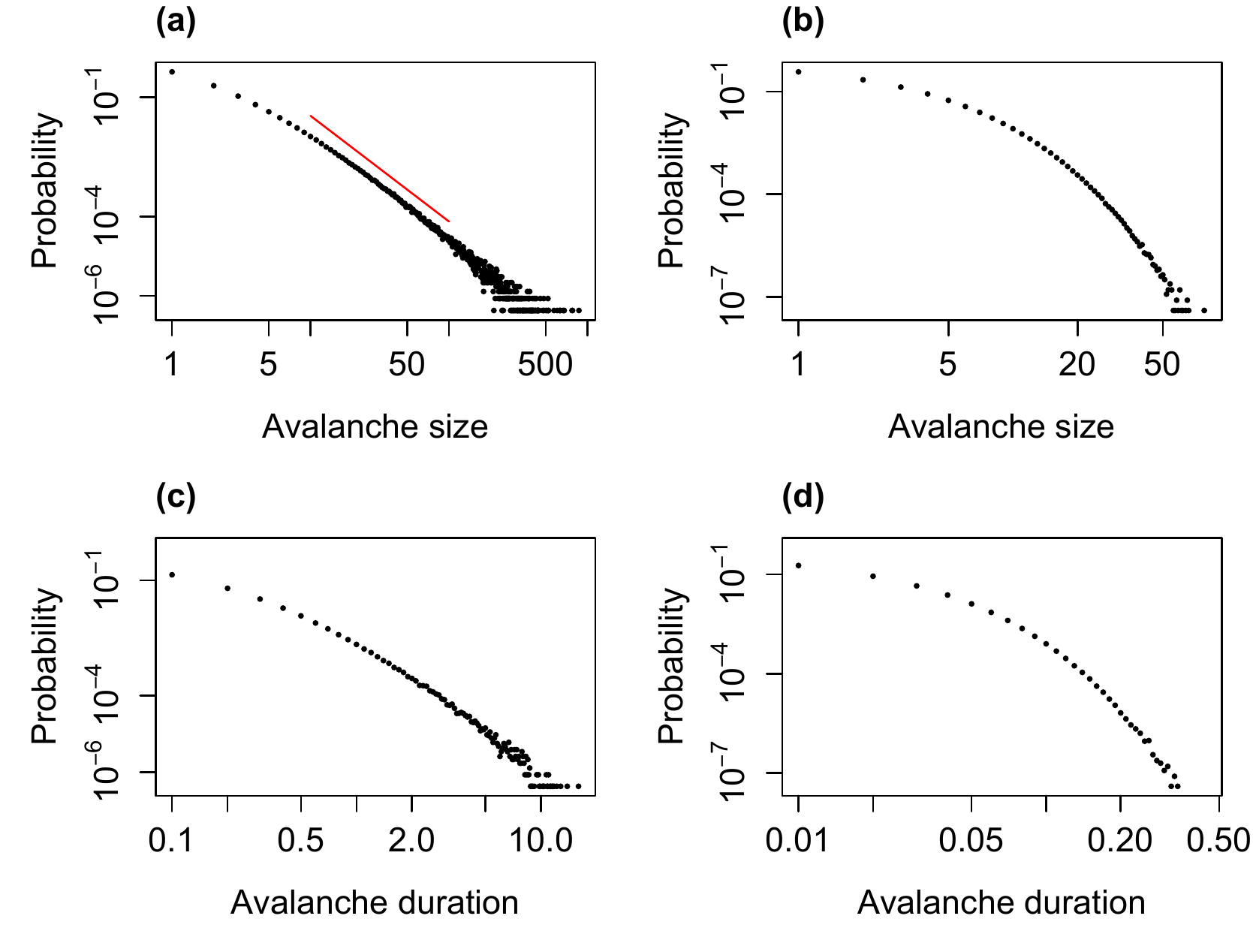}
\caption{\textbf{Distributions of avalanche size and duration for the system driven from subcritical and supercritical regimes.} Avalanche size and duration distributions for the system driven from (a,c) subcritical ($\alpha =1.1$) and (b,d) supercritical ($\alpha=0.9$) states. Simulations were run with $N=800,\ w=1,\ h=1/N$. The red line in (a) shows the linear fit with a slope of -2.66.}
\label{fig:subsupsizedists}
\end{figure}

\section*{Appendix 3: Altering the activation function}

Throughout this paper we considered a linear activation function. What happens if a different activation function is chosen? Do we observe the same type of dynamics? In this appendix we briefly investigate two other activation functions: an exponential and a quadratic.

First let us consider the system with an exponential activation function such that:
\[
\frac{dA}{dt}=\left(\frac{1}{1+e^{-(\frac{w}{N}A+h)}}-\frac{1}{2}\right)(N-A)-\alpha A.
\]
This function saturates and so is somewhat more realistic than the linear function considered previously. Also, note that the function is set such that when $A$ and $h$ are both zero we also have $f(x)=0$, i.e.\ without any external input and with no active neurons the network will remain in this state.
With this activation function the eigenvalues of the fixed points are given by:
\[
\lambda = f'(x)(N-A)-f(x)-\alpha.
\]
We have that
\[
f'(x)=\frac{\frac{w}{N}e^{-(\frac{w}{N}A+h)}}{(1+e^{-(\frac{w}{N}A+h)})^{2}}=\frac{w}{N}\left(f(x)+\frac{1}{2}\right)\left(\frac{1}{2}-f(x)\right)=\frac{w}{N}\left(\frac{1}{4}-f^{2}(x)\right),
\]
and so this could be used to find a critical fixed point along with the fact that at the fixed point of the system we have that:
\[
f(x)=\frac{\alpha A}{(N-A)},
\]
which defines the level of the external input at the critical fixed point.

As before, consider initially the case where there is no external input ($h=0$). In this case $A=0$ is a fixed point, which is critical (with $\lambda=0$) if and only if $\alpha =w/4$ by the above equations. What happens to this system in the presence of small external input? Fig.~\ref{fig:saturatingraster} shows the raster plot for the three different levels of the external input considered previously: $h=0.01/N,\ 0.1/N,\ 1/N$. Comparing with Fig.~\ref{fig:examplesims}, the firing rate is lower with the saturating function studied here, however, the overall pattern of firing is the same. For all three levels of the external input we continue to observe avalanche dynamics and for lower levels of the external input (as the system approaches the critical regime) these avalanches become more distinct. Fig.~\ref{fig:saturatingloglog} shows the avalanche size, duration and IAI distributions for each of these three levels of the external input. Comparing with Figures~\ref{fig:distsize}~and~\ref{fig:IAIdists} we find that a similar relationship with the critical regime emerges. With $h=1/N$ the IAI distribution shows scale-free behaviour (note that by the same derivation as previously, theoretically the distribution is a weighted sum of hypoexponentials). For lower levels of the external input the scale-free behaviour in the IAI distribution is lost but the distribution of avalanche sizes appears scale-free. As was shown previously for the system with a linear activation function, we also found that when $h=1/N$ the IAIs exhibited LRTCs up to a crossover point (data not shown). For lower levels of the external input these correlations were lost.

\begin{figure}
\centering
	\includegraphics[width=150mm]{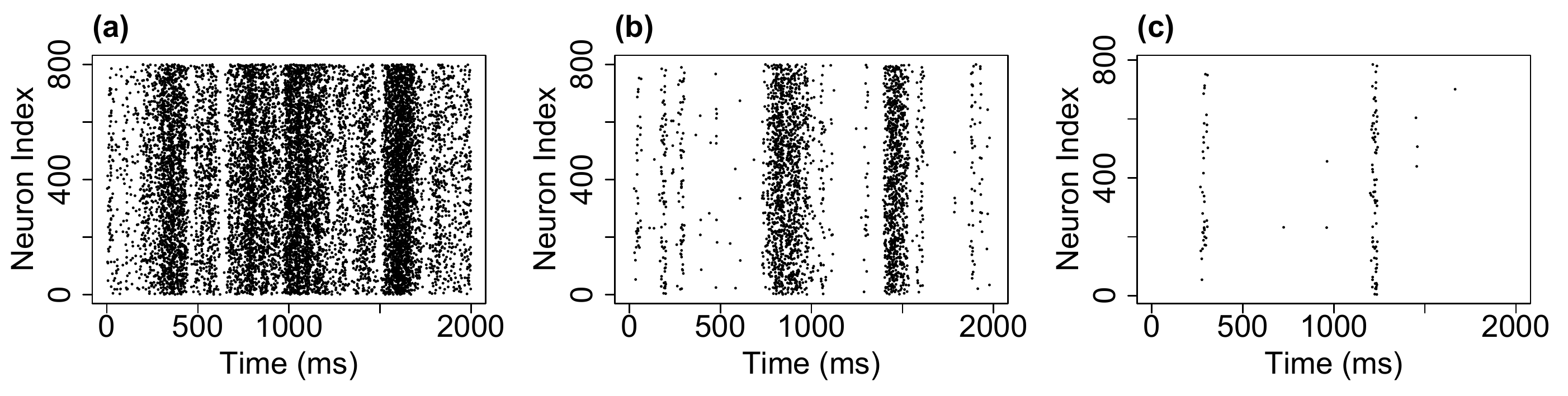}
\caption{\textbf{Raster plots for different levels of the external input with the saturating activation function.} Neuronal firing during the first 2 seconds of example simulations with (a) $h=1/N$, (b) $h=0.1/N$ and (c) $h=0.01/N$. For all three simulations $w=1, \alpha=0.25$ and $N=800$. Comparing with Figure~\ref{fig:examplesims} we find that while in this case the firing rate is lower (note the longer time scale over which the raster plot is displayed) the overall pattern is the same, with the avalanches becoming more distinct with lower levels of the external input.}
\label{fig:saturatingraster}
\end{figure}

\begin{figure}
\centering
	\includegraphics[width=150mm]{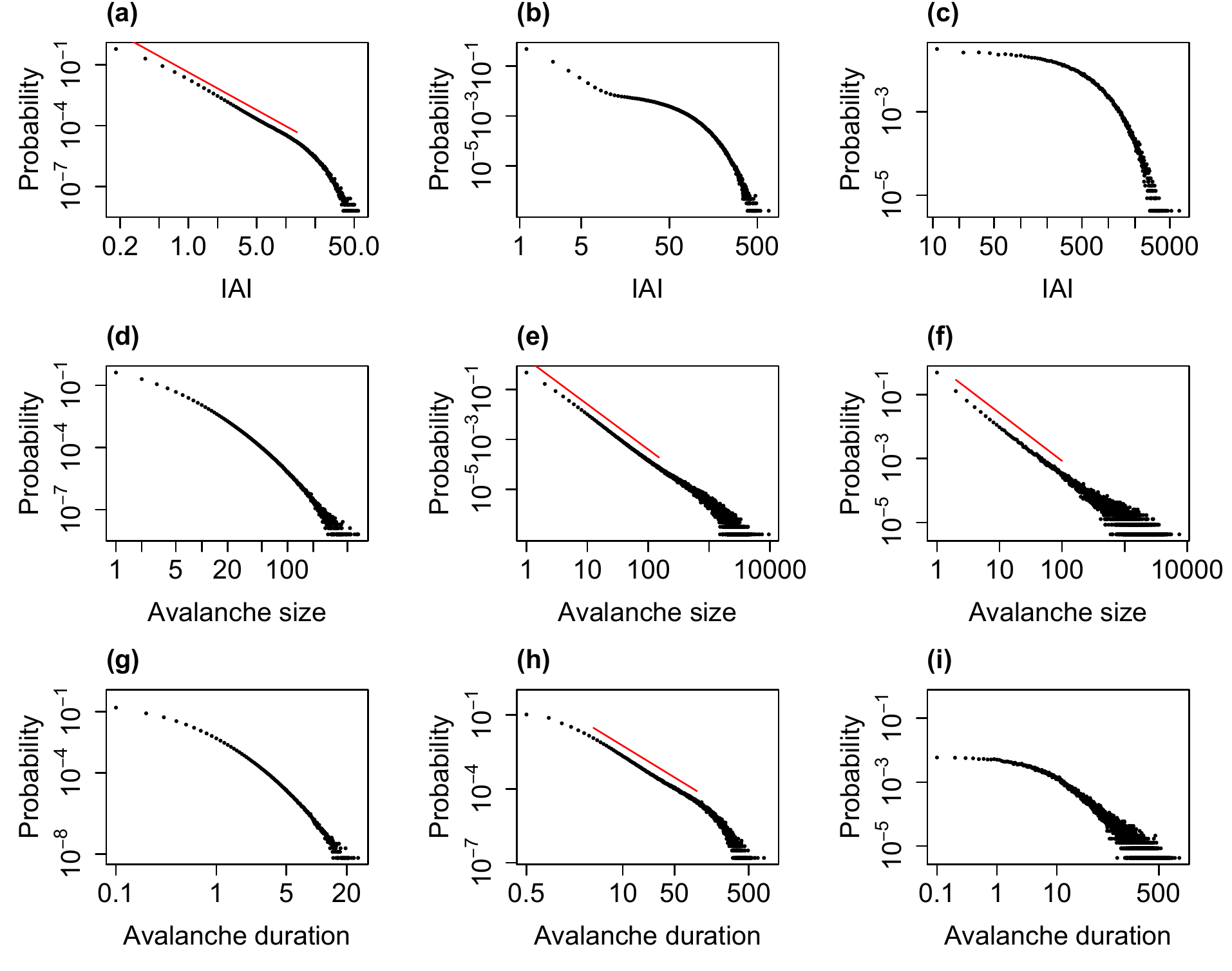}
\caption{\textbf{IAI, avalanche size and duration distributions for the system with the saturating activation function.} The distributions are for simulations with $h=1/N$ (a,d,g), $h=0.1/N$ (b,e,h) and $h=0.01/N$ (c,f,i). For all simulations $N=800, \alpha =0.25, w=1$ and the distributions are pooled from 10 simulations each of length $10^4$ seconds. The red lines indicates linear fits on the double logarithmic scale, i.e.\ fitted power-laws with exponents of (a) 2.65, (e) 1.81, (f) 1.49 and (h) 1.84.}
\label{fig:saturatingloglog}
\end{figure}

With both the linear and saturating activation function we considered the system in the region of the critical regime, with the system driven from the critical regime by the positive external input. Consider the system instead with a quadratic activation function:
\[
\frac{dA}{dt}=\left( \frac{w}{N}A^{2}+h\right) (N-A) -\alpha A,
\]
With this activation function the fixed points are given by:
\[
g(A)=\frac{dA}{dt}=-\frac{w}{N}+wA^{2}-(h+\alpha )A+hN=0,
\]
with eigenvalues:
\[
\lambda=g'(A)=-3\frac{w}{N}A^{2}+2wA-h-\alpha.
\]
Solving these simultaneously we find:
\[
\alpha = -3\frac{w}{N}A^{2}+2wA-h,
\]
\[
2wA^{3}-wNA^{2}+hN^{2}=0,
\]
which define the parameter space and the value of the fixed point for which a critical fixed point can be obtained. Thus, we find that unlike the model with the linear (and saturating) activation function, here with a non-zero external input it is possible to tune the system so that it is directly at the critical regime.

Upon examining this parameter space one can note that in many cases there also exists a stable (positive) fixed point as well as the critical fixed point. From simulating such a system we found (data not shown) that the dynamics of the system are quickly attracted to the stable fixed point and so the critical fixed point has little affect on the dynamics. Therefore, to have a system which is affected by a critical fixed point in the presence of a non-zero external input (in the case of this activation function and where positive parameters are required) the critical regime must be the only fixed point of the system. Given that $g(A)$ is a cubic equation, to achieve a single fixed point which is critical this point must be an inflection point with $g'(A)=0$ and $g''(A)=0$.  From these equalities we find that the critical fixed point is $A=N/3$ and we must also have $h=\frac{wN}{27}$ and $\alpha = \frac{8wN}{27}$.

Fig.~\ref{fig:quadratic} shows a raster firing plot and the number of active neurons throughout a simulation for the system with a single critical fixed point. As would be expected, the number of active neurons fluctuates about the critical point. Previously when considering avalanche dynamics we have binned the firing. However, as noted in the discussion the binning method will always separate firing into avalanches and as there are no clear periods of inactivity this does not seem appropriate here. Recall that in the zero input case (see the companion paper) we seeded the system so perturbing it away from the fully quiescent state (which was the critical fixed point) and defined an avalanche as the firing that occurred before the system returned to the fully quiescent state. In a similar approach here it is possible to define an avalanche as the number of neurons that fire in a single excursion from the critical fixed point. We therefore counted the number of neurons that fired from when the system was deflected (either in a positive or negative direction) from the fixed point ($A=N/3$) until the next time at which the system had exactly $N/3$ active neurons. Fig.~\ref{fig:quadratic} shows the probability distribution of the size of the avalanches defined in this way. The distribution appears to be scale-free over a range of scales. 

Thus, while critical dynamics may not be apparent initially when examining data (for example if we were to look at the overall dynamics from the simulations with quadratic activation function), we can observe signatures of criticality when the dynamics are examined in relation to the known critical regime.  Here we can note that the network firing fluctuates about the critical regime - that is the number of active neurons fluctuates about this regime and so the average number of active neurons across the course of a simulation is approximately equal to the critical state of $N/3$. It might therefore be interesting to examine the fluctuations about the mean activity level in experimental settings where activity is continuous (i.e.\ cannot be described as intermittent avalanche-like activity) to determine whether signatures of criticality are present. Indeed, such an approach has been taken previously to examine MEG data, thresholding at the median level~\cite{poil2008}.

\begin{figure}
\centering
	\includegraphics[width=150mm]{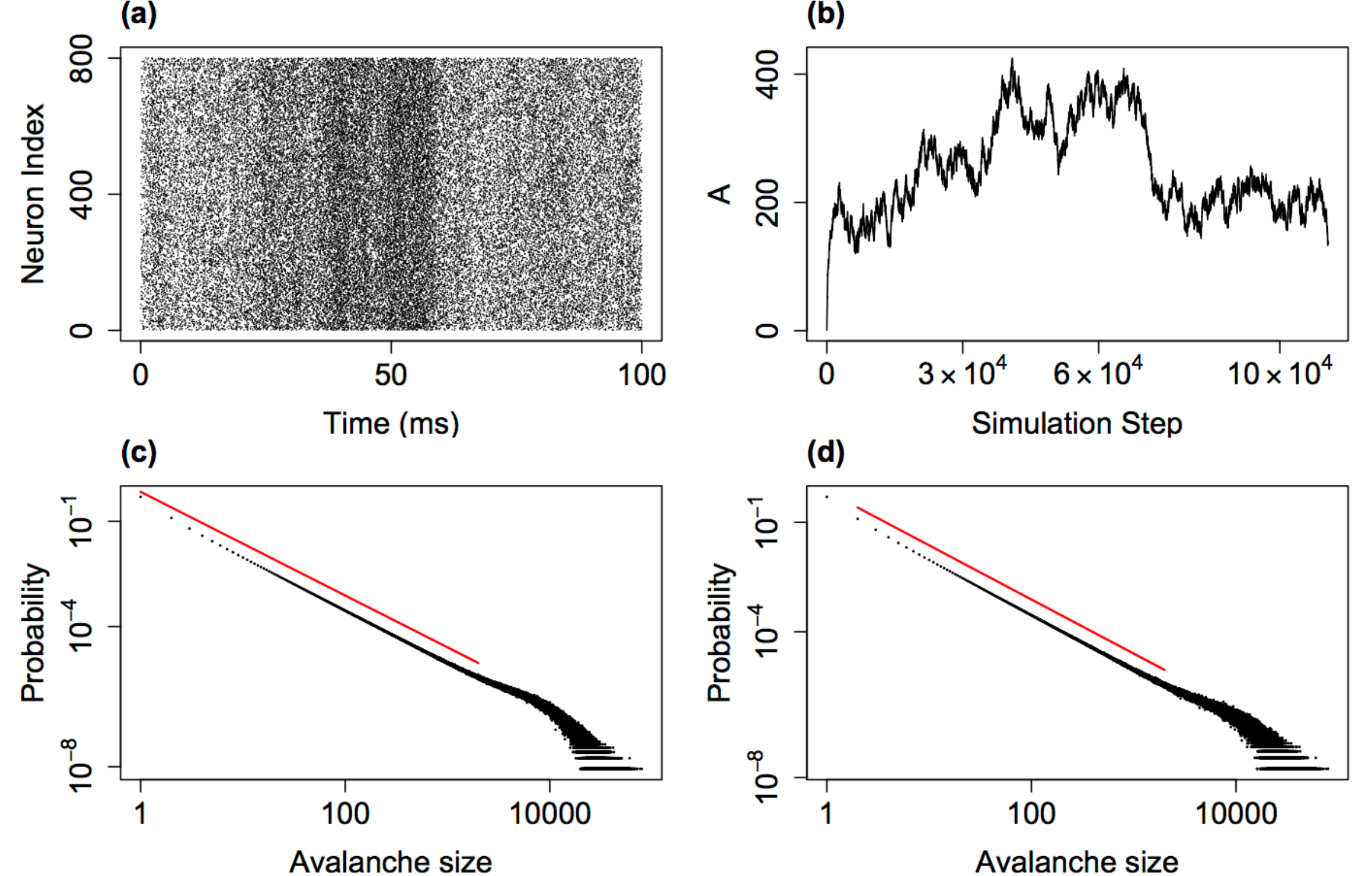}
\caption{\textbf{Dynamics of the network with a quadratic activation function.} (a) Raster plot of network firing for simulation with $N=800,\ h=\frac{wN}{27},\ \alpha=\frac{8wN}{27},\ w=0.01$. (b) For the same simulation this plot shows the number of active neurons (A) at each simulation step. The distribution of avalanches (c) and positive avalanches only (d) - as described in the main text - pooled from 20 simulations of 1000 seconds in length. The red lines indicate linear fits, i.e. fitted power-laws, both with exponents of 1.48.}
\label{fig:quadratic}
\end{figure}

\bigskip

%%%%%%%%%%%%%%%%%%%%%%%%%%%%%%%%
%\section*{Author's contributions}
%    Text for this section \ldots

%%%%%%%%%%%%%%%%%%%%%%%%%%%
\section*{Acknowledgements}
  \ifthenelse{\boolean{publ}}{\small}{}
   Caroline Hartley is funded through CoMPLEX (Centre for Mathematics and Physics in the Life Sciences and Experimental Biology), University College London. Timothy Taylor is funded by a PGR studentship from MRC, and the Departments of Informatics and Mathematics at University of Sussex. Istvan Z. Kiss acknowledges support from EPSRC (EP/H001085/1). Simon Farmer acknowledges support from the National Institute for Health Research University College London Hospitals Biomedical Research Centre.
 
%%  Bmc_article.bst  will be used to                       %%
%%  create a .BBL file for submission, which includes      %%
%%  XML structured for BMC.                                %%
%%  After submission of the .TEX file,                     %%
%%  you will be prompted to submit your .BBL file.         %%

\newpage
{\ifthenelse{\boolean{publ}}{\footnotesize}{\small}
 \bibliographystyle{bmc_article}  % Style BST file
  \bibliography{AQ_nzh.bib} }     % Bibliography file (usually '*.bib' ) 

%%%%%%%%%%%

\ifthenelse{\boolean{publ}}{\end{multicols}}{}

\end{bmcformat}
\end{document}